\documentclass[floatfix,twocolumn,showpacs,preprintnumbers,amsmath,amssymb,prb]{revtex4}
\usepackage{amsmath}
\usepackage{bm}
\usepackage{graphics}
\usepackage{dcolumn}
\begin{document}
\title{Spin-spin coupling in electrostatically coupled quantum dots}
\author{Mircea Trif, Vitaly N. Golovach, and Daniel Loss}
\affiliation{Department of Physics and Astronomy, University of Basel, Klingelbergstrasse 82, CH-4056 Basel, Switzerland}
\date{\today}

\begin{abstract}
We study the spin-spin coupling between two single-electron quantum dots due to the Coulomb 
and spin-orbit interactions, in the absence of tunneling between the dots. 
We find an anisotropic $XY$ spin-spin interaction that is proportional to the Zeeman splitting 
produced by the external magnetic field.
This interaction is studied both in the limit of weak and strong Coulomb repulsion with respect to the
level spacing of the dot.
The interaction is found to be a non-monotonic function of inter-dot distance $a_0$ and external magnetic field,
and, moreover, vanishes for some special values of $a_0$ and/or magnetic field orientation. 
This mechanism thus provides a new way to generate and tune spin interaction between quantum dots.
We propose  a scheme to measure this spin-spin interaction based on the spin-relaxation-measurement technique.
\end{abstract}

\date{\today}
\draft{}
\preprint{}
\maketitle

\section{Introduction}
\label{Introlabel}
%The electron spin in semiconductor devices has a large potential for applications 
%in spintronics\cite{Prinz,Wolf,ALS} and quantum computing\cite{LDV,IAB,ALS,Coish}.
%The need for a scalable system that is both well-decoupled from the environment, 
%and hence quantum coherent, and available for manipulation by classical 
%fields (gates) is compromised in the electron spin in semiconductor quantum 
%dots\cite{LDV}.
%The transverse spin lifetime is extremely long in GaAs quantum dots
%and extends to $100\;{\rm ms}$ at low magnetic fields ($\sim 1\;{\rm T}$).
%Long spin coherence times ($T_2^*\sim(1-100)\;{\rm ns}$) have been first observed
%for extended electrons in ZnCdSe quantum wells\cite{KSSA} and $n$-type GaAs bulk 
%semiconductors\cite{KA}.

The spin degree of freedom of electrons in semiconductor nanostructures has attracted much interest recently.
Besides its applications in spintronics\cite{ALS}, which makes use of the spin in the same 
way as the charge in conventional devices,
there has  been considerable interest in the quantum coherence of spin in different materials and geometries at the nanoscale \cite{Coish}. 
The long coherence times measured some time ago in semiconductors 
 \cite{KA} supported the idea of quantum computing with spins localized in quantum dots  \cite{LDV,IAB}--a rapidly growing field now, where  recent experimental progress has been remarkable \cite{Ciorga,Ono,Elzerman,Kroutvar,Petta,Imamoglu,Koppens,Zumbuhl}.
 
Electron spins in semiconductor nanostructures, however,  are not decoupled from the charge degree of freedom, one of the primary 
reasons  for this coupling being the spin-orbit interaction. This coupling  leads to many interesting 
phenomena in 
the physics of semiconductors \cite{ALS,Coish}, an important one being the ability to control the spin 
of the electron 
with electric fields acting on its charge degrees of freedom \cite{MR,Dobrowolska,Salis,RE2,DL,GBL}. In quantum dots the spin-orbit interaction manifests itself as a weak perturbation 
when compared with the confinement energy. The measure of smallness is given by the ratio between the dot radius $\lambda$ and the 
spin-orbit length, $\lambda_{SO}$---the distance over which an electron travels and thereby precesses by an angle 
$\pi$ about the intrinsic 'magnetic' field induced by the spin-orbit interaction. 
Despite its smallness the spin-orbit interaction is very important for the coherence of the spin dynamics. 
For example, spin-orbit interaction allows for coupling of the electron spin to (bosonic) environments, 
such as phonons \cite{KN,GKL} or particle-hole excitations in quantum point contacts \cite{BGL}, which in turn causes relaxation and decoherence of the spins. 
%However, spin-orbit coupling can also induce coherent interactions. 
%Bosonic baths, being governed by quantum mechanics themselves, can lead to coherent spin-spin %interactions among 
%spatially separated electron spins via the vacuum fluctuations \cite{IAB,privman}. 
Moreover, electrons being charged particles interact via the long range Coulomb forces 
with each other,
even if they are confined 
to well-separated quantum dots with no overlap of their wave functions.
%As a consequence, the modifications in electron charge is translated via spin-orbit interaction into spin state modification. 
Through this electrostatic coupling and in combination with the spin orbit interaction, the spins of two electrons located in different  dots become coupled even in the absence of 
tunneling between the dots. 

In this paper we provide a detailed analysis of such an effective spin-spin interaction
for lateral quantum dots in a configuration as shown in Fig. 1.
We will see
that the origin of this interaction is the ``tidal'' effect each of the electrons produces on the charge distribution
of the other electron via  electrostatic forces.
Because of the spin-orbit interaction, the electric dipole moment (as well as higher moments) 
in each dot couples to its electron spin.
As a result, the two spins experience an interaction resembling a magnetic dipole-dipole interaction\cite{Abragam} with  effective magnetic moments which can be strongly enhanced by up to a factor of $10^3$ compared to the Bohr magneton.
%While the magnetic dipolar interaction is due to the electron magnetic moment,
%the spin-spin interaction studied in this paper can be related, at the leading order, to a spin-electric moment 
%that is produced by a combination of the spin-orbit and Zeeman interactions.
The magnitude of the spin-spin coupling obtained via this spin-electric effect 
can  be efficiently controlled and even completely suppressed by adjusting external parameters such as the magnetic 
field direction, strength, and inter-dot distance. 
%While we focus our study on lateral quantum dots, the
%system could also consist of two vertical quantum dots very close to each 
%other, but without tunneling between them.

The spin-spin interaction can, in principle, be used to perform two-qubit operations as required in the spin-based quantum 
computing scheme\cite{LDV}, because it entangles spins and can easily be switched on and off.
We note that a similar mechanism for spin interaction based on electrostatic coupling
was studied very recently in Ref.~\onlinecite{chinezii} 
for vertically coupled quantum dots, and in Ref.~\onlinecite{FL} for the special case of one-dimensional 
quantum dots formed in semiconducting nanowires. 
Besides the differences in geometry and dimensionality, both of these works treat only the case of weak Coulomb interaction (compared to the level spacing), 
while we treat here also the opposite limit of strong Coulomb interaction where new and interesting features emerge. 
In the limiting case of strongly elliptical dots we recover the one-dimensional results obtained in Ref.~\onlinecite{FL}.

We emphasize again that in the present study we exclude tunneling and thus the type of spin interaction studied in the following is fundamentally different from the Heisenberg exchange interaction for which  the presence of electron tunneling between the 
dots is crucial\cite{BLV,hu:2000a,schliemann:2001a,note1}. Similarly, the combined effect of  Heisenberg exchange interaction
and spin-orbit coupling\cite{KVK,BSV,DBVBL,SB,BuL} is also based on tunneling and should be
carefully distinguished from the spin-orbit effect studied here. 
We also note that the Heisenberg exchange coupling allows typically for much stronger
spin-spin coupling than the electrostatically induced one. For instance,  in GaAs dots the Heisenberg exchange
can reach values  on the order of $ 0.1$ meV $-1$ meV, which, as we shall see, exceeds the
electrostatically induced spin coupling by three to four orders of magnitude. 
Nevertheless, the electrostatic spin coupling can prove useful for cases where it is difficult to
get sufficient wave function overlap (needed for large Heisenberg exchange), and, moreover, it is also important to understand the electrostatic spin-effects in detail in order to get control over possible interference effects
between different types of spin coupling. This will be for example of importance for spin qubit applications in order  to minimize spin decoherence and gate errors.
%We stress that the present type of spin-spin interaction has a different origin as the pure exchange one, 
%which comes as a consequence of spin modification  during the virtual tunneling and consequently changing the usual isotropic Heisenberg-type of interaction. This type of coupling via spin-orbit interaction was studied by several authors \cite{KVK,BSV,DBVBL,SB,BuL}. 

Finally, in view of experimental tests we propose a scheme to measure the spin-spin interaction in a double dot setup with a nearby charge detector.
We propose to combine the spin-measurement technique of Ref.~\onlinecite{Elzerman} with the
entangling property of the spin-spin interaction and present a gate pulsing sequence
that enables one to access the coupling constant in the time domain by measuring
the occupation probability of a Zeeman sublevel.

The paper is organized as follows. In Sec.~\ref{themodellabel} 
we introduce our model for the  quantum-dot system  and its basic Hamiltonian. 
In Sec.~\ref{sscouplinglabel} we derive the effective spin-spin coupling Hamiltonian within our model. 
In this section we analyze in detail the two limiting cases of  weak and strong Coulomb 
interaction, and obtain various forms of the spin coupling  as function
of magnetic field strength and orientation, as well as of the interdot distance. 
The most general form of the spin coupling  tensor is given in the Appendix.
In Sec.~\ref{measurementschemelabel}, we propose a scheme to measure the spin-spin coupling.
There, we also discuss the effect of the hyperfine interaction with the nuclear spins
on the measurement scheme.
Finally, in Sec.~\ref{discconclabel} we discuss our results and end with some conclusions.

\section{The Model}
\label{themodellabel}
Our system consists  of two electrons each of which is localized in a quantum dot, and the two dots are separated from each other, without tunneling between them. The system  is composed of two gate-defined quantum dots in a two-dimensional semiconductor layer ($e.g.$ GaAs or InAs). A schematics of the system we consider is shown in Fig 1.

\begin{figure}[t]
\scalebox{0.5}{\includegraphics{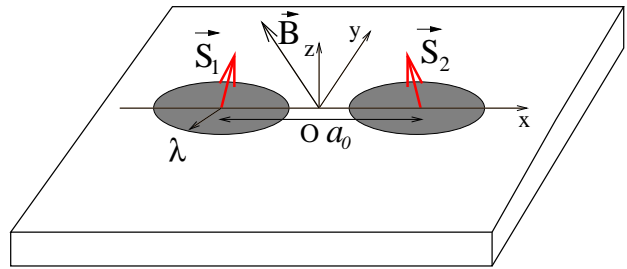}}
\caption{The figure shows a sketch  of the model  system which consists of two identical quantum dots in the $xy$-plane, separated by  distance $a_0$ (measured from dot-center to dot-center). $\vec{S}_i$ denotes the spin of electron $i=1,2$, $\lambda$ is the dot radius, and $\vec{B}$ is the external magnetic field.
The respective orbital wave functions of electron 1 and 2 are assumed to have no overlap
(i.e. tunneling between the dots is excluded). The remaining purely electrostatic
Coulomb interaction between the electron charges leads, via spin-orbit interaction, to
an effective coupling between their spins. This spin-spin interaction depends sensitively
on the  orientation  of $\vec{B}$, with no component along it, and is proportional to $\vec{B}^{2}$. }
\label{Fig1}
\end{figure}

We model the system by a harmonic confinement potential, which, for simplicity is assumed to be the same for both dots. 
Each dot is assumed to contain one electron with charge $-e$ and spin 
$\mbox{\boldmath $S$}=(\hbar/2)\bm{\sigma}$, 
with $\bm{\sigma}=(\sigma_x,\sigma_y,\sigma_z)$ being the Pauli matrices. 
The model Hamiltonian consisting  of several terms  reads
\begin{equation}
H  =  H_0+H_Z+H_C+H_{SO}, \label{1}
\end{equation}
where $H_0$ is the energy of the two electrons in the confinement potentials
\begin{equation}
H_0 = \sum_{i=1,2}\left(\frac{p^2_i}{2m^*}+U(\bm{r}_i)\right). \label{2}
\end{equation}
Here, $\bm{p}_i=-i\hbar\partial/\partial\bm{r}_i+(e/c)\bm{A}(\bm{r}_i)$ is the 2D kinetic momentum of the $i$-th electron at position $\bm{r}_i$,  $m^*$  the effective mass, $c$  the speed of light,  $U(\bm{r}_i)=(m^*/2)\omega_0^2r_i^2$  the confinement potential for the $i$-th electron which  is assumed to be harmonic and isotropic, and $\bm{A}(\bm{r}_i)$ is the electromagnetic vector potential. The strength of the confinement energy is given by the frequency $\omega_0$.  
The second term on the right-hand side of Eq. (\ref{1}) is the Zeeman energy of the two
electrons,
\begin{equation}
H_Z  =  \frac{1}{2}g\mu_B\bm{B}\cdot(\bm{\sigma}_1+\bm{\sigma}_2). \label{3}
\end{equation}
The third term in Eq. ($\ref{1}$) is the unscreened Coulomb interaction between the two electrons,
\begin{equation}
H_C  =  \frac{e^2}{\kappa|\bm{r}_1-\bm{r}_2+\bm{a}_0|}, \label{4}
\end{equation}
where $\kappa$ is the dielectric constant of the material and $a_0$ is the geometric distance between  the two dots, namely between the potential minima ('center') of the dots. With this choice, we measure the distance for each electron from its own dot center.  The last term in Eq. ($\ref{1}$)
is the spin-orbit coupling  which for strong
$z$-confinement is given by
\begin{equation}
H_{SO}  =  \sum_{i=1,2}[\beta(-p_x^i\sigma_x^i+p_y^i\sigma_y^i)+\alpha(p_x^i\sigma_y^i-p_y^i\sigma_x^i)], \label{5}
\end{equation}
being the sum of the Dresselhaus term \cite{Dress} ($\beta$)  coming from bulk inversion asymmetry and the Rashba term \cite{Rashba} ($\alpha$) coming from
structure inversion asymmetry. We assume the
same coefficients $\beta$ and $\alpha$ for both dots. 
It is convenient to work with center-of-mass and relative coordinates \cite{JP}, as the Coulomb
interaction couples only to the relative  ones  and  the solution of the center-of-mass part is straightforward \cite{Fock,Darwin}. This then involves the standard substitutions  $M=2m^*$, $m=m^*/2$, $\bm{R}=(\bm{r}_1+\bm{r}_2)/2$,  $\bm{r}=\bm{r}_1-\bm{r}_2$, and    $\bm{P}=\bm{p}_1+\bm{p}_2$ and $\bm{p}=(\bm{p}_1-\bm{p}_2)/2$.

\section{Spin-Spin Coupling}
\label{sscouplinglabel}
We now turn our attention  to the spin-orbit interaction. As  was shown in Ref. \onlinecite{GKL}, the spin-orbit coupling gives  non-zero first order effects only if a magnetic field is present, as  a consequence of the Kramers degeneracy. In order to describe the effective first order spin-orbit term in the presence of a magnetic field we make use   of the  Schrieffer-Wolff (unitary) transformation\cite{GKL,GBL}
\begin{equation}
e^S(H_d+H_{SO})e^{-S}=H_d+\Delta{H},
\end{equation}
where $S=-S^{\dagger}$ is chosen such that $P\Delta{H}=\Delta{H}$, with the projector operator $P$ satisfying $PA=\sum_n A_{nn}|n\rangle\langle n|$ $\forall A$, and $H_d|n\rangle=E_n|n\rangle$. The Hamiltonian $H_d=H_0+H_C$ (or $H_d=H_R+H_r+H_C$ in center-of-mass and relative coordinates).  The Hamiltonian $\tilde{H}=H_d+\Delta{H}$ is diagonal in the basis of $H_d$ and has the same energy spectrum as the Hamiltonian $H=H_d+H_{SO}$. In first order of the spin-orbit interaction $H_{SO}$ the transformation generator  becomes $S  =  (1-P)L^{-1}_d H_{SO}$, 
%\begin{equation}
%S  =  (1-P)L^{-1}_d H_{SO}, \label{6}
%\end{equation}
where $L_d$ is the dot Liouvillean, $L_d A=[H_R+H_r+H_C,A]$, $\forall A$. Evaluating this expression 
explicitly we obtain  
\begin{equation}
S=(1-P)\sum_{i=1,2}\bm{\xi}_{i}\cdot \bm{\sigma}_{i}, \label{6}
\end{equation}
with $\bm{\xi}_{1,2}=(y_{1,2}/\lambda_+,x_{1,2}/\lambda_-,0)$. In this lowest  order the transformed Hamiltonian $\tilde{H}$ becomes
\begin{equation}
\tilde{H} =  H_R+H_r+H_Z+H_C+H_{SO}^Z, \label{7}
\end{equation}
with $H_{SO}^Z=H_{SO}^{Z 1}+H_{SO}^{Z 2}$ where
\begin{equation}
H_{SO}^{Z 1,2}=[S,H_Z^{1,2}] = E_Z(1-P)[\bm{l}\times(\bm{\xi}_R\pm\bm{\xi}_r/2)]\cdot\bm{\sigma}_{1,2}. \label{8}
\end{equation}
In Eq. ($\ref{8}$) $\bm{l}=\bm{B}/B$ is the magnetic field direction vector, $E_Z=g\mu_BB$ is the Zeeman energy and the vectors $\bm{\xi}_R$ and $\bm{\xi}_r$ are given  by $\bm{\xi}_R=(Y'/\lambda_-,X'/\lambda_+,0)$ and $\bm{\xi}_r=(y'/\lambda_-,x'/\lambda_+,0)$, respectively. The new coordinates correspond to a rotation by an angle $\pi/4-\gamma$ with respect to the coordinate frame in which the direction of the $\bm{a}_0$-vector is associated with the $x$-axis in the $XY(xy)$ plane so that the final expressions have the simplest form \cite{GKL}. Here,  $\gamma$ is the angle between the $xy$ frame  in Fig. 1 and the normal axes of the crystal. The spin-orbit lengths $\lambda_{\pm}$ are given in the form $1/\lambda_{\pm}=m^*(\beta\pm\alpha)$. We are now in a position  to derive the coupling between the two spins. This is achieved by performing a second Schrieffer-Wolff (SW2) transformation which excludes the first order contribution in spin-orbit interaction with no diagonal matrix elements. The new transformed Hamiltonian has the form $H_{{\it eff}}=e^T\tilde{H}e^{-T}$, with $T$ given by
\begin{equation}
T  =  (1-P)(L_d+L_Z)^{-1}H_{SO}^{Z}. \label{9}
\end{equation}
We assume now that the Zeeman energy is smaller than the orbital confining energy, $E_Z\ll\hbar\omega_0$, which is usually the case for electrons in quantum dots, such that we can neglect in Eq. ($\ref{9}$) the Zeeman Liouvillean $L_Z$ (for spin-orbit effects due to level crossing see Ref. \onlinecite{BL}). In second-order in spin-orbit coupling the effective Hamiltonian $H_{{\it eff}}$ becomes
\begin{equation}
H_{{\it eff}}  =  H_d+H_Z+\frac{1}{2}[L_d^{-1}H_{SO}^Z,H_{SO}^Z]. \label{10}
\end{equation}
The last term in Eq. ($\ref{10}$) contains the desired spin-spin coupling between the two spins. However, besides this interaction it also contains some self-interaction terms which renormalizes only the Zeeman splitting. We will not study those terms since they are of no practical interest in the case of identical dots. We consider a general magnetic field $\bm{B}=(\cos{\Phi}\sin{\theta},\sin{\Phi}\sin{\theta},\cos{\theta})$, where $\theta$ is the angle between the magnetic field and the $z$-axis perpendicular to the 2DEG plane and $\Phi$ the angle  between the in-plane component of the magnetic field and  the  $x$-direction (Fig. 1). The interaction between the two  spins has the most general form
\begin{equation}
\tilde{H}_s  =  \frac{1}{2}\sum_{i\neq j}[L_d^{-1}H_{SO}^{ Z i},H_{SO}^{ Z j}], \quad  i,j=1,2.\label{11}
\end{equation} 
The spin Hamiltonian is obtained  by  averaging over the orbital ground state,  $H_s=\langle 0|\tilde{H}_s|0\rangle$. We then obtain 
\begin{equation}
H_s=\bm{\sigma}_1\cdot\overline{\overline{M}}\bm{\sigma}_2,
\end{equation}
where
\begin{equation}
\overline{\overline{M}}_{ab}=E_Z^2\langle 0|[(\bm{l}\times L_d^{-1}\bm{\xi}_1)_a,(\bm{l}\times\bm{\xi}_2)_b]|0\rangle,\;\; a,b=x,y,z.
\end{equation}
We note that there is no component of  the spin along the magnetic field direction as a consequence of the vector product in the tensor $\overline{\overline{M}}$. By diagonalizing the above tensor, we obtain for the Hamiltonian $H_s$ the reduced expression 
\begin{equation}
H_s  =  J_{\tilde{x}}\sigma_{\tilde{x}}^1\sigma_{\tilde{x}}^2+J_{\tilde{y}}\sigma_{\tilde{y}}^1\sigma_{\tilde{y}}^2. \label{12}
\end{equation}
%The expressions for $J_{\tilde{x},\tilde{y}}$ can be expressed in the form
%\begin{equation}
%J_{\tilde{x},\tilde{y}}=\frac{E_Z^2}{2}\left(A+B\pm\sqrt{(A+B)^2-4C^2}\right),
%\end{equation}
where the couplings $J_{\tilde{x},\tilde{y}}$ depend on the magnetic field orientation and on the functions $C_{a_1b_2}=\langle 0|[L_d^{-1}a_1,b_2]|0\rangle$, with $a,b=x,y$ (for explicit expressions see the Appendix). Thus, the effective spin-spin interaction is highly anisotropic, and, in general, of the $XY$-type. We note in particular that for an in-plane magnetic field  ($\theta=\pi/2$), the spin Hamiltonian reduces to the Ising Hamiltonian, $H_s=J_{\tilde{y}}\sigma_{\tilde{y}}^1\sigma_{\tilde{y}}^2$ (in a transverse magnetic field). 
Next, we  rewrite  $H_s$  in terms of raising/lowering spin operators $\sigma_{\pm}=\sigma_{\tilde{x}}\pm i\sigma_{\tilde{y}}$
\begin{equation}
H_s=J_{{\it eff}}(\sigma_{+}^1\sigma_-^2+\sigma_-^1\sigma_+^2)+ J'_{{\it eff}}(\sigma_{-}^1\sigma_-^2+\sigma_+^1\sigma_+^2),\label{100}
\end{equation}
with $J_{{\it eff}}=(1/2)(J_{\tilde{x}}+J_{\tilde{y}})$ and $J'_{{\it eff}}=(1/2)(J_{\tilde{x}}-J_{\tilde{y}})$. 
We recall now that the \textit{ full} spin Hamiltonian includes the Zeeman energy, given in Eq. (\ref{3}),
which leads to a large energy gap with $2E_Z\gg J_{\tilde{x},\tilde{y}}$. We will find below that typically 
\begin{equation}
\frac{J_{\tilde{x},\tilde{y}}}{E_Z}\sim \frac{E_Z}{\hbar\omega_0}\left(\frac{\lambda}{\lambda_{SO}}\right)^2\ll 1
\end{equation}
under our assumption that $E_Z\ll\hbar\omega_0$ and $\lambda\ll\lambda_{SO}$.
As a consequence,  we can neglect in Eq. ($\ref{100}$) the terms proportional to $J'_{{\it eff}}$ since they cause transitions between different Zeeman levels of the total spin. The  relevant spin-spin interaction, $H^{{\it eff}}_s$, which acts only within the $S-T_0$ subspace, becomes then
\begin{equation}
H_s^{{\it eff}} = J_{{\it eff}}(\sigma_{+}^1\sigma_-^2+\sigma_-^1\sigma_+^2)\label{13}.
\end{equation}
Thus, we are left with the task of calculating the coupling strengths $J_{\tilde{x},\tilde{y}}$ and  $J_{{\it eff}}$.  Because of the Coulomb term, Eq. ($\ref{4}$), this cannot be done exactly and some approximations are required. They will depend on the ratio $\delta$ between the  Coulomb interaction strength, $e^2/\kappa a_0$, and the orbital level spacing, $\hbar\omega_0$, giving $\delta=(e^2/\kappa a_0)/\hbar\omega_0=(\lambda/a_B)\cdot(\lambda/a_0)$, with $\lambda=\sqrt{\hbar/m^*\omega_0}$ being  the dot radius and $a_B=\hbar^2\kappa/m^*e^2$ - the Bohr radius in the material. In other words, the  parameter $\delta$ will dictate the physics of the system, and from now on we will speak of the ratio $\lambda/a_B$ as being the Coulomb interaction strength (representing in fact the 'true' Coulomb strength for touching dots). For making the following analysis more transparent we focus on  the case with only  Rashba spin-orbit coupling 
($\lambda_-=\lambda_+\equiv\lambda_{SO}$). The generalization to the case with both Rashba and Dresselhaus terms  present is straightforward, but at the cost of more complicated expressions (see Appendix).

\subsection{Weak Coulomb coupling\;-\;$\delta\ll 1$}
\label{weakCoulombcouplabel} 
One interesting case is met when $\delta\ll 1$, such that the Coulomb interaction can be treated as a perturbation compared to the orbital level spacing. In this case, one can retain only the first order contribution from the Coulomb interaction, which translates into the approximation  $L^{-1}_d \approx L_0^{-1}-L_0^{-1}L_CL_0^{-1}$. Making use of this and after some algebra we obtain for the spin-spin coupling the following expression
\begin{equation}
H_s  = \int d\bm{r}_1d\bm{r}_2\frac{\delta{\rho}_1\delta{\rho_2}}{\kappa|\bm{r}_1-\bm{r}_2+\bm{a}_0|},\label{14}
\end{equation}
where the operators $\delta{\rho_i}$, $i$=1,2, are the charge density distribution modifications in each dot as a consequence of the spin-orbit interaction. They  are defined as
\begin{equation}
\delta{\rho_i}  =  \rho_i-\rho_i^0, \quad i=1,2, \label{15}
\end{equation}
with $\rho_i^0$ being the charge density operator in the absence of spin-orbit interaction and $\rho_i=e^{T_0}\rho_i^0e^{-T_0}$ the one in the presence of spin-orbit interaction, with $T_0=L_0^{-1}H_{SO}^Z$ for the present approximation. From  Eq. ($\ref{14}$) we see that the spin interaction results from a Coulomb-type of coupling between two charge density distributions which themselves depend on spin. 

Let us now analyze in more detail  Eq. ($\ref{14}$). The first task is to find $\delta\rho_i$, for $i=1,2$, namely the spin-orbit induced charge distribution or the spin-dependent charge distributions for each dot. In order to do this, we give first some important relations valid in the case of harmonic confining potential, relations which are used in the following for the derivation of the main results
\begin{equation}
L_0^{-1}x_i=-\frac{i}{\hbar m^*\omega_0^2}\left(p_x^i+\frac{eB_z}{c}y_i\right)\label{50}
\end{equation}
\begin{equation}
L_0^{-1}y_i=-\frac{i}{\hbar m^*\omega_0^2}\left(p_y^i-\frac{eB_z}{c}x_i\right)\label{51}
\end{equation}
\begin{equation}
L_0^{-1}\bm{p}_i=\frac{im^*}{\hbar}\bm{r}_i.\label{52}
\end{equation}
 Making use of the relations Eqs. ($\ref{50}$-$\ref{52}$) and  within the first order of spin-orbit coupling, $i.e.$ $\delta\rho_i\approx[T_0,\rho_{i}^0]$, we obtain
\begin{eqnarray}
\delta\rho_i(\bm{r})=\frac{2E_Z e}{m^*\lambda^{2}\omega_0^2\lambda_{SO}}\,\rho_{0i}\big[\cos{\theta}(y_i\cos{\Phi}+x_i\sin{\Phi})\sigma_x^i\nonumber\\+(y_i\sin{\Phi}-x_i\cos{\Phi})\sigma_y^i\big],\;\;\;\;\;\;\;\;\;\;\;\;\label{53}
\end{eqnarray}
with $\rho_{i}^0$ being the bare charge density in the dot corresponding to the ground state and which assumes the well-known form for harmonic potentials
\begin{equation}
\rho_{i}^0(\bm{r})=\frac{1}{\pi\lambda^{2}}e^{\displaystyle{-\frac{(x_i^2+y_i^2)}{\lambda^{2}}}}.\label{54}
\end{equation}
We note that when there exist a  perpendicular component of the magnetic field, the dot radius is renormalized due to the orbital effect of the magnetic field  $\lambda\rightarrow\lambda/\sqrt{1+r^2}$, with $r=\omega_0/2\omega_c$ ($\omega_c=eB_z/m^*c$, $B_z=B\cos{\theta}$). However, we will still refer to $\lambda$ as being the dot radius, with the appropriate expression depending on the magnetic field orientation. We could now insert the expression Eq. ($\ref{53}$) for $\delta\rho_i$ in Eq. ($\ref{14}$) and compute directly the spin Hamiltonian. However, working with the Coulomb potential, it is more convenient to work with the center-of-mass and relative coordinates and for simplicity the $x$-axis along the inter-dot direction $\bm{a}_0$. Assuming for simplicity a perpendicular magnetic field, the spin Hamiltonian $H_s$ takes the form
\begin{eqnarray}
H_s=\frac{4E_Z^2 e^2}{m^{*2}\lambda^{4}\omega_0^4\lambda_{SO}^2}\iint d\bm{r}\,d\bm{R}\,\rho_0{(r)}\rho_0{(R)}\nonumber \\\times \frac{(X^2-x^2/4)\sigma_x^1\sigma_x^2+(Y^2-y^2/4)\sigma_y^1\sigma_y^2}{\kappa\sqrt{y^2+(x+a_0)^2}},\label{55}
\end{eqnarray}
with the electronic densities $\rho_0{(r)}=(2/\pi\lambda^{2})\exp(-r^2/2\lambda^{2})$ and $\rho_0{(R)}=(1/2\pi\lambda^{2})\exp(-2R^2/\lambda^{2})$. In Eq. ($\ref{55}$) there are no mixed terms like $\sigma_x^1\sigma_y^2$ since those terms vanish because of the odd symmetry of the integrands in the case of harmonic confinement, which reflects inversion symmetry. The limit of in-plane magnetic field is obtained very easy from Eq. ($\ref{55}$) by substituting the denominator with $[(X^2-x^2/4)\cos^2{\Phi}+(Y^2-x^2/4)\sin^2{\Phi}]\sigma_{\tilde{y}}^1\sigma_{\tilde{y}}^2$. [For general  field orientation the expression for $H_s$ is more complicated (see Appendix).]  In order to make the following analysis more transparent, we introduce the dimensionless coordinates $\bm{r}\rightarrow \bm{r}/\lambda$ and $\bm{R}\rightarrow \bm{R}/\lambda$. The integration over the center-of-mass coordinates is now straightforward and the reduced expression for the spin Hamiltonian becomes
\begin{equation}
H_s=\frac{E_Z^2}{m^{*2}\lambda\omega_0^4\lambda_{SO}^2}\left(\Delta E_C^x\sigma_x^1\sigma_x^2+\Delta E_C^y\sigma_y^1\sigma_y^2\right),\label{56}
\end{equation}
for a perpendicular magnetic field and
\begin{equation}
H_s=\frac{E_Z^2}{m^{*2}\lambda\omega_0^4\lambda_{SO}^2}\left(\Delta E_C^x\sin^2{\Phi}+\Delta E_C^y\cos^2{\Phi}\right)\sigma_{\tilde{y}}^1\sigma_{\tilde{y}}^2,\label{56b}
\end{equation}
for an in-plane magnetic field oriented at an angle $\Phi$ with respect to the inter-dot distance vector $\bm{a}_0$. The energy differences $\Delta E_C^{x,y}$ are given by
\begin{equation}
\Delta E_C^x=\frac{e^2}{\kappa\lambda^2}\int d\bm{r}\rho_0(r)\frac{1-x^2}{\sqrt{y^2+(x+a_0/\lambda)^2}},\label{57}
\end{equation}
\begin{equation}
\Delta E_C^y=\frac{e^2}{\kappa\lambda^2}\int d\bm{r}\rho_0(r)\frac{1-y^2}{\sqrt{y^2+(x+a_0/\lambda)^2}}.\label{58}
\end{equation}
The ground state and the first excited states of the dots in relative coordinates give rise to different charge distributions ($\rho_0$, $\rho_{1x}$, and $\rho_{1y}$, respectively), and thus to different potential energies seen by a test charge at a distance $a_0$ (along $x$) away from the center of the charge distribution (in  relative coordinates). $\Delta E_{C}^{x,y}$ are the differences between these potential energies.

 Before studying the distance dependence of the spin Hamiltonian $H_s$ (determined by $\Delta E_C^{x,y}$) in the entire range of distances, it
 is instructive to see how  the expression Eq. ($\ref{14}$)  behaves in
 the large distance limit, $a_0\gg\lambda$, and to make some comparison with the magnetic 
 dipolar interaction in  vacuum \cite{Coish}. We perform a
 multipole expansion of the Hamiltonian in Eq. ($\ref{14}$). The first
 non-zero contribution takes the form of a dipole-dipole interaction
 between two spin-dependent electric dipoles, or phrased differently,
 the interaction between two charge-induced magnetic dipoles
\begin{equation}
H_s\approx\frac{\bm{m}_1\cdot\bm{m}_2-3(\bm{m}_1\cdot\bm{n}_a)(\bm{m}_2\cdot\bm{n}_a)}{\kappa\,a_0^3},\label{dipole1}
\end{equation}
with the dipole moments $\bm{m}_i$ given by
\begin{equation}
\bm{m}_i=Tr_{orb}[\delta\rho_i\bm{r}_i]=\bar{\bar{\mu}}\,\bm{\sigma}_i\;,i=1,2.
\end{equation}
Here, the trace is taken over the orbital degrees
of freedom with $\bm{n}_a=\bm{a}_0/a_0$ and  $\bar{\bar{\mu}}$ being
the tensor corresponding to an effective spin-orbit induced 
magneton
\begin{eqnarray}
\bar{\bar{\mu}}=\frac{eE_Z}{m^*\omega_0^2\lambda_{SO}}\left(
\begin{matrix}
-\cos{\theta} & 0 & 0 \cr 0 & \cos{\theta} & 0 \cr 0 & \sin{\theta} & 0
\end{matrix}\right).
\label{magneton}
\end{eqnarray}  
We see from Eq. ($\ref{magneton}$) that the tensor $\bar{\bar{\mu}}$ depends on the magnetic field orientation with respect to the 2DEG and that it is also anisotropic, in contrast to the usual isotropic 
Bohr magneton $\mu_B=e\hbar/2m_ec$ ($m_e$ is the mass of the free electron and $c$ the speed of light). We note that the $z$-component of the induced magnetic moment (with $\hat{\bm{z}}||\bm{B}$) vanishes, $i.e.$ $\bm{m}=(m_x,m_y,0)$. Let us quantify the strength of $\bar{\bar{\mu}}$ by the norm  $||\bar{\bar{\mu}}||=(1/3)\sqrt{\sum_{i,j}\mu^2_{ij}}$, $i.e.$ 
\begin{equation}
||\bar{\bar{\mu}}||=\frac{1}{3}\frac{eE_Z}{m^*\omega_0^2\lambda_{SO}}\sqrt{1+\cos^2{\theta}}.
\end{equation}
We compare now   $||\bar{\bar{\mu}}||$ with $\mu_B$. First of all, we note that
    $||\bar{\bar{\mu}}||$ vanishes when there is no Zeeman
    splitting. However, for finite magnetic fields,
    $||\bar{\bar{\mu}}||$ can exceed $\mu_B$ by many orders of magnitude in
    the case of quantum dots. To give an estimate, we assume
    $\hbar\omega_0\approx 0.5\,{\rm meV}$, $E_Z\approx 0.05\, {\rm
    meV}$ ($B\approx 2\, {\rm T}$)  and $m^*=0.067 m_e$,
    $\lambda_{SO}\approx 10^{-6}\,{\rm m}$ for GaAs quantum dots. With
    these values, and taking $\theta=0$ (perpendicular magnetic field) we obtain
\begin{equation}
\frac{||\bar{\bar{\mu}}||}{\mu_B}=\frac{4}{3}\,\frac{E_Z}{\hbar\omega_0}\,\frac{m_e}{m^*}\,\frac{c}{\omega_0\lambda_{SO}}\approx 10^3.
\end{equation}
 
We describe now in more detail  the limit
of large distance between the dots. From Eqs. ($\ref{56}$) and ($\ref{55}$) we find for $a_0\gg\lambda$
\begin{equation}
H_s=J(\sigma_y^1\sigma_y^2-2\sigma_x^1\sigma_x^2),\label{57}
\end{equation}
for a perpendicular magnetic field, and
\begin{equation}
H_s=J(\cos^2{\Phi}-2\sin^2{\Phi})\sigma_{\tilde{y}}^1\sigma_{\tilde{y}}^2
\end{equation}
for an in-plane magnetic field, with the coupling strength $J$ having the form
\begin{equation}
J=\frac{E_Z^2\,e^2}{\kappa m^{*2}\omega_0^4\lambda_{SO}^2a_0^3}.\label{66}
\end{equation}
From Eq. ($\ref{66}$) we see a large distance decay $\sim a_0^{-3}$, $i.e.$ a long range type behavior.
%with the $x$ component of the interaction being twice the corresponding $y$ component and negative. This is because in the large distance limit the $x$ component has a lower electrostatic energy in the excited state than 
%that in the ground state, while for the $y$ component it is the opposite. 
We note also that the large distance result in Eq. ($\ref{66}$) does not depend anymore on the orbital effect of the 
magnetic field. 
Working instead with the effective Hamiltonian defined in Eq. ($\ref{13}$), the effective coupling strength $J_{{\it eff}}$ for arbitrary magnetic field
is given by
\begin{equation}
J_{{\it eff}}=-\frac{J}{2}\left[1+\sin^2{\theta}(1-3\sin^2{\Phi})\right].\label{ferro}
\end{equation} 
 We note that the spin coupling can range from maximally ferromagnetic interaction with $J_{{\it eff}} =-J$
  at $\theta=\pi/2$ and $\Phi=0$ to maximally antiferromagnetic interaction with $J_{{\it eff}} =J/2$
  at $\theta=\pi/2$ and $\Phi=\pi/2$, and with $J_{{\it eff}}$ passing through zero for certain angles.

\begin{figure}[t]
\scalebox{0.35}{\includegraphics{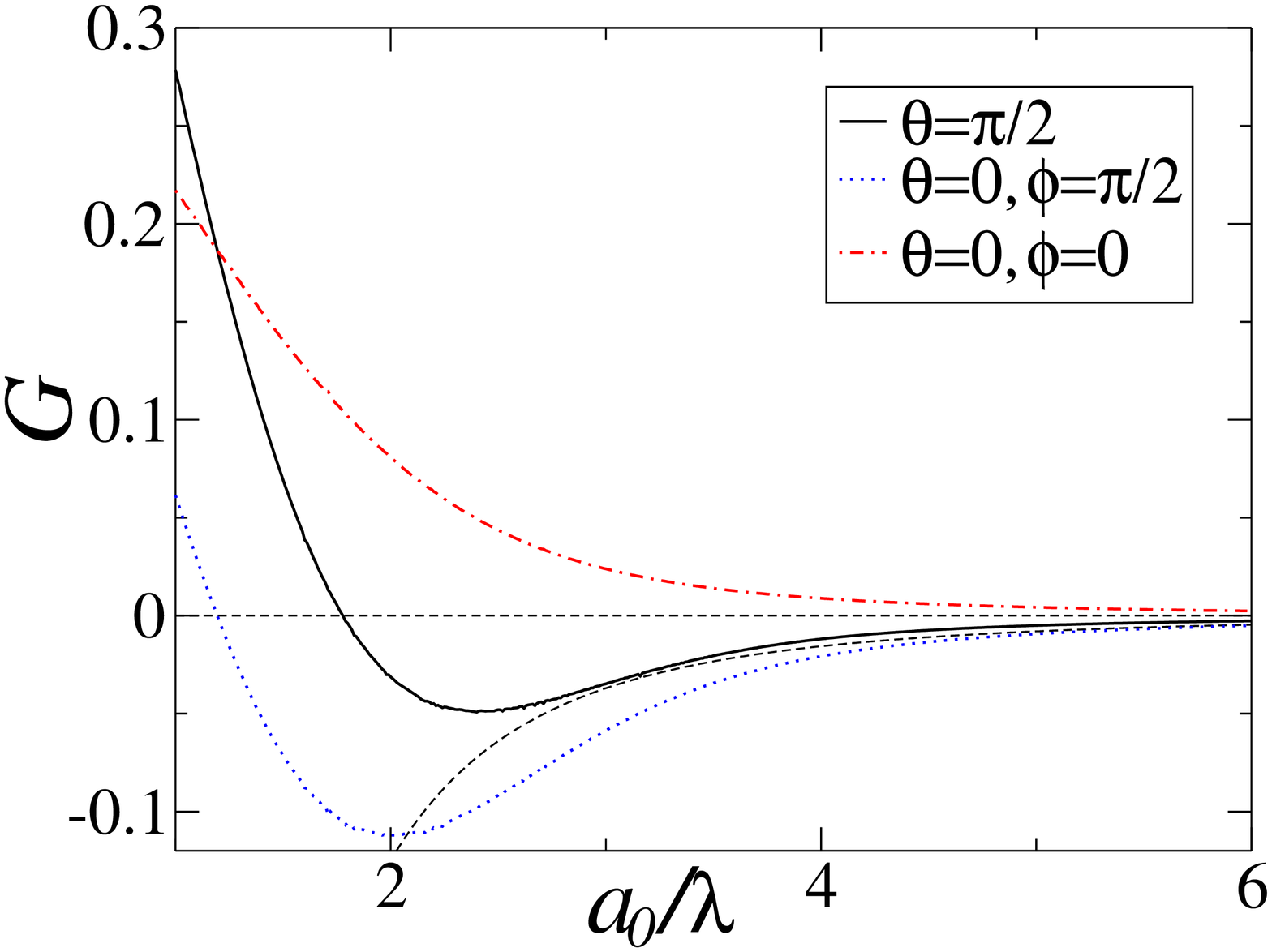}}
\caption{The function $G$ occurring  in Eq. ($\ref{fa}$) plotted  as a function of the geometric 
distance $a_0$ between the dot centers scaled by the dot radius $\lambda$ for different magnetic field orientations. 
The dashed line represents the dipolar approximation of $G$ for a perpendicular magnetic field 
($\theta=0$) which scales like $a_0^{-3}$.}
\label{Fig2}
\end{figure}

Next, we consider the case of arbitrary distance but still with small Coulomb interaction strength as defined at the 
beginning of this section. 
Considering $H^{{\it eff}}_s$ defined in Eq. ($\ref{13}$) we obtain 
\begin{equation}
J_{{\it eff}}=E_Z\,\frac{\lambda}{a_B}\,\frac{E_Z}{\hbar\omega_0}\,\left(\frac{\lambda}{\lambda_{SO}}\right)^2G(a_0/\lambda,\theta,\Phi), \label{16}
\end{equation}
where
\begin{eqnarray}
G(a_0/\lambda,\theta,\Phi)=\frac{\kappa\lambda}{e}\big[(\cos^2{\theta}\cos^2{\Phi}+\sin^2{\Phi})\Delta E_C^y\nonumber\\+(\cos^2{\theta}\sin^2{\Phi}+\cos^2{\Phi})\Delta E_C^x\big].\;\;\;\;\;\;\;\;\label{fa}
\end{eqnarray}
The function $G(a_0/\lambda,\theta,\Phi)$  is plotted in Fig. 2 for different angles $\theta$ and $\Phi$. As for the  large distance limit in Eq. ($\ref{ferro}$), a similar but more complicated ferromagnetic-antiferromagnetic crossover behavior occurs as a function of the field orientation. However, in this case this behavior can also be induced  by changing the distance between the dots $a_0$ (see Fig. \ref{Fig2}). 

 Eq. ($\ref{16}$) suggests that the condition $\delta\ll1$ is too restrictive. Instead, the weaker condition $(\lambda/a_B)(\lambda/a_0)^3\ll1$ is sufficient for the approximation to be valid.
Fig. 2 shows a breakdown of the dipolar approximation (i.e. of the $a_0^{-3}$ decay), occurring at a dot separation $a_0/\lambda\approx2$ for perpendicular magnetic fields ($\theta=0$), and also a cancellation of 
this interaction for some given distance, which is around  $a_0/\lambda\approx1.8$. 
This shows that the sum of the two electrostatic energy differences $\Delta E_C^x+\Delta E_C^y$ has a non-monotonic 
behavior as a function of the distance $a_0$. 
Actually, only $\Delta E_C^x$ is non-monotonic, whereas $\Delta E_C^y$ has a positive value which decreases with $a_0$, 
as can be seen from Fig. 2. 
If an in-plane magnetic field is applied along $y$ ($\Phi=0$) or $x$ ($\Phi=\pi/2$) direction, 
we obtain a dependence only either on $\Delta E_C^y$ or on $\Delta E_C^x$. 
Accordingly, $G(a_0/\lambda)$ will be larger  in some parameter range 
as compared with the case of perpendicular fields, see Fig. 2. 
% We note that the function $F(a_0/\lambda)$ depends only on the ratio $a_0/\lambda$ 
% and the sole restriction being $\delta\ll 1$ in order for this theory to be valid.
At this point  it is instructive to consider numerical estimates for the coupling strength $J_{{\it eff}}$. 
For this we consider GaAs quantum dots for which we assume $\lambda_{SO}\approx \;10^{-6}\; {\rm m}$, $m^*=0.067\,m_e$, 
$g=-0.44$, $\kappa=13$, and also consider $\hbar\omega_0\approx0.5\; {\rm meV}$, $E_Z=0.05\; {\rm meV}$ 
($B\approx 2\; {\rm T}$) and $a_0=5\cdot10^{-7}\; {\rm m}$. 
These estimations lead to a coupling strength $J\approx\,10^{-10}\;{\rm eV}$, which lead to a time dynamics 
of the order of $10^{-5}\;{\rm s}$. 
If this time scale is  longer than the decoherence times in GaAs quantum dots, the system will be 
insensitive to the coherent dynamics induced by the coupling  $J_{{\it eff}}$. 
Shorter time scales are obtained for materials with larger spin-orbit coupling
such as InAs.
The spin-orbit length $\lambda_{SO}$ in this material is comparable with a typical  dot 
size of about $100\, {\rm nm}$. 
Even though our perturbative approach starts to get unreliable in this case, it still can provide 
a rough estimate for the coupling strength. For InAs we have $m^*=0.023m_e$, $g=14.8$, $\kappa=13$, $\lambda_{SO}\approx 100 {\rm nm}$, 
and we choose $\hbar\omega_0\approx 1 {\rm meV}$, $E_Z\approx 0.1 {\rm meV}$ and $a_0/\lambda\approx 3$. 
With those values we obtain for the coupling $J_{{\it eff}}\approx 10^{-7} {\rm eV}$, 
which corresponds to a switching time of about $\sim 50 {\rm ns}$ for a swap of the spin states
of electron one and electron two.
This time scale for the spin dynamics is  shorter than the expected spin decoherence time in such quantum dots. 
Thus, this interaction mechanism provides a useful way for the dynamical control of the spin-spin coupling. As discussed before, for an in-plane magnetic field the coupling constant could even be higher, depending on the angle of the magnetic field with respect to the inter-dot axis.

\subsection{Elliptical dots with $\delta\ll1$}
\label{ellipticaldotslabel}
We briefly generalize  the previous results to elliptical dot shapes. This will also allow us
to study the one-dimensional limit and recover previous results obtained for
one-dimensional nanowires\cite{FL}.
We  consider  elliptical dots which are characterized by the frequencies $\omega_{0x}$ and $\omega_{0y}$ corresponding to the $x$ and $y$ directions, respectively. In this case,  Eq. ($\ref{66}$) is replaced by
\begin{equation}
H_s=\frac{e^2E_Z^2}{m^{*2}\lambda_{SO}^2}\left(\frac{\Delta{E}_C^x}{\omega_{0x}^4\lambda_1^2}\sigma_x^2\sigma_x^2+\frac{\Delta{E}_C^y}{\omega_{0y}^4\lambda_2^2}\sigma_y^1\sigma_y^2\right),\label{elliptic}
\end{equation}
where the electrostatic energies $\Delta{E}_C^{x,y}$ become now
\begin{equation}
\Delta{E}_C^x=\frac{1}{\kappa\lambda_1}\int d\bm{r}\rho_0(\bm{r})\frac{1-x^2}{\sqrt{y^2+(x+a_0/\lambda_1)^2}},
\end{equation}
\begin{equation}
\Delta{E}_C^y=\frac{1}{\kappa\lambda_2}\int d\bm{r}\rho_0(\bm{r})\frac{1-y^2}{\sqrt{y^2+(x+a_0/\lambda_2)^2}},
\end{equation}
with the charge density distribution function, expressed in  relative coordinates, 
\begin{equation}
\rho_0(\bm{r})=\frac{2}{\pi\lambda_1\lambda_2}e^{\displaystyle{-x^2/2\lambda_1^2-y^2/2\lambda_2^2}}.
\end{equation}
For perpendicular magnetic fields and elliptical dots the lengths $\lambda_{1,2}$ are given  by \cite{Cederbaum}
\begin{equation}
\lambda_{1,2}=\sqrt{\frac{4\hbar(n+1)}{m_{1,2}^*\omega_{1,2}}},
\end{equation}
where $n=m_1m_2\omega_1\omega_2\omega_c^2/B^2$, $\omega_{1,2}=\sqrt{A\pm B}/2$, and $m_{1,2}=2B/(C\pm\omega_c^2+B)$ with the explicit expressions for $A$, $B$, and $C$ 
\begin{equation}
A=\omega_{0x}^2+\omega_{0y}^2+\omega_c^2,
\end{equation}
\begin{equation}
B=\sqrt{(\omega_{0x}^2+\omega_{0y}^2+\omega_c^2)^2-4\omega_{0x}^2\omega_{0y}^2},
\end{equation}
\begin{equation}
C=\omega_{0x}^2-\omega_{0y}^2.
\end{equation}
Taking now also the limit of strongly elliptical dots, $i.e.$ $\omega_{0y}\gg\omega_{0x},\omega_c$, we see that this is equivalent to keeping only one component of the spin interaction, namely the $\sigma_x^1\sigma_x^2$ part, and that the orbital effect of the magnetic field drops out. The resulting
Hamiltonian then becomes
\begin{equation}
H_s=\frac{e^2E_Z^2\Delta{E}_C^x}{m^{*2}\omega_{0x}^4\lambda^2\lambda_{SO}^2}\sigma_x^1\sigma_x^2+O\big((\omega_{0x}/\omega_{0y})^4\big).
\end{equation}
Considering now the large distance limit, $a_0\gg\lambda$, analogously  to Eq. ($\ref{57}$), our result reduces formally to the one in Ref. \onlinecite{FL}, $i.e.$
\begin{equation}
H_s=-2\frac{E_Z^2\,e^2}{\kappa m^{*2}\omega_{0x}^4\lambda_{SO}^2a_0^3}\sigma_x^1\sigma_x^2+O\big((\omega_{0x}/\omega_{0y})^4\big).\label{FL}
\end{equation}
The above expression,  Eq. ($\ref{FL}$), could be obtained directly from Eq. ($57$) since within the
considered limit  there is no orbital effect of the perpendicular magnetic field on the spin-spin interaction. We note that in this limit the resulting spin-spin coupling takes the form of an Ising interaction which, together with single qubit rotations,
can be used\cite{LDV}  to   efficiently perform CNOT gate operations between two qubits. We finally note  that in one dimensions the Rashba interaction can be treated exactly, leading to a renormalization of the $g$-factor\cite{FL}, $g\rightarrow g\exp{(-\lambda^2/\lambda_{SO}^2)}$. This exact treatment is no longer possible in the 2D case considered here, except for the special case\cite{SEL} when  $\alpha=\pm\beta$ and the problem becomes effectively 1D\cite{SEL}.

\subsection{Strong Coulomb coupling\;-\;$\delta\ge1$}
\label{strongCoulombCoupllabel}
We turn now to the more involved case of strong  Coulomb interaction strength, $\delta\ge1$, which cannot be treated  perturbatively. However, some approximations are still possible and we will explore two of them in the following section. The first approximation consists in reducing the two electron system to two classical point-charge particles.  The classical equilibrium condition will be obtained by minimizing the total potential energy of the two particles. By doing this, the motion of the electrons will take place around the new equilibrium positions obtained from the equation
\begin{equation}
a^2(a-a_0)=2\lambda^4/a_B, \label{19}
\end{equation}
where $a_0$ is the initial geometric distance and $a$ the effective distance between the electrons in classical equilibrium. However, we are interested in the motion around the equilibrium position, which means that for  small deviations, we may substitute the full Coulomb interaction with an  effective one, remembering that $\bm{r}=\bm{r}_1-\bm{r}_2$,
\begin{equation}
\frac{e^2}{\kappa |\bm{r}+\bm{a}_0|} \rightarrow  \frac{e^2}{2\,\kappa\,a^3}\,(3(\bm{n}_a\cdot\bm{r})^2-r^2).\label{20}
\end{equation}     
We note  that the coordinates are measured now with respect to the new equilibrium position.  Within this approximation, the relative  Hamiltonian $H_r$ is replaced with the  new, renormalized one, $\widetilde{H}_r$
\begin{equation}
\widetilde{H}_r  =  \frac{p^2}{2m}+\frac{1}{2}m\omega_x^2x^2+\frac{1}{2}m\omega_y^2y^2,\label{21}
\end{equation}
\begin{figure}[t!]
\scalebox{0.4}{\includegraphics{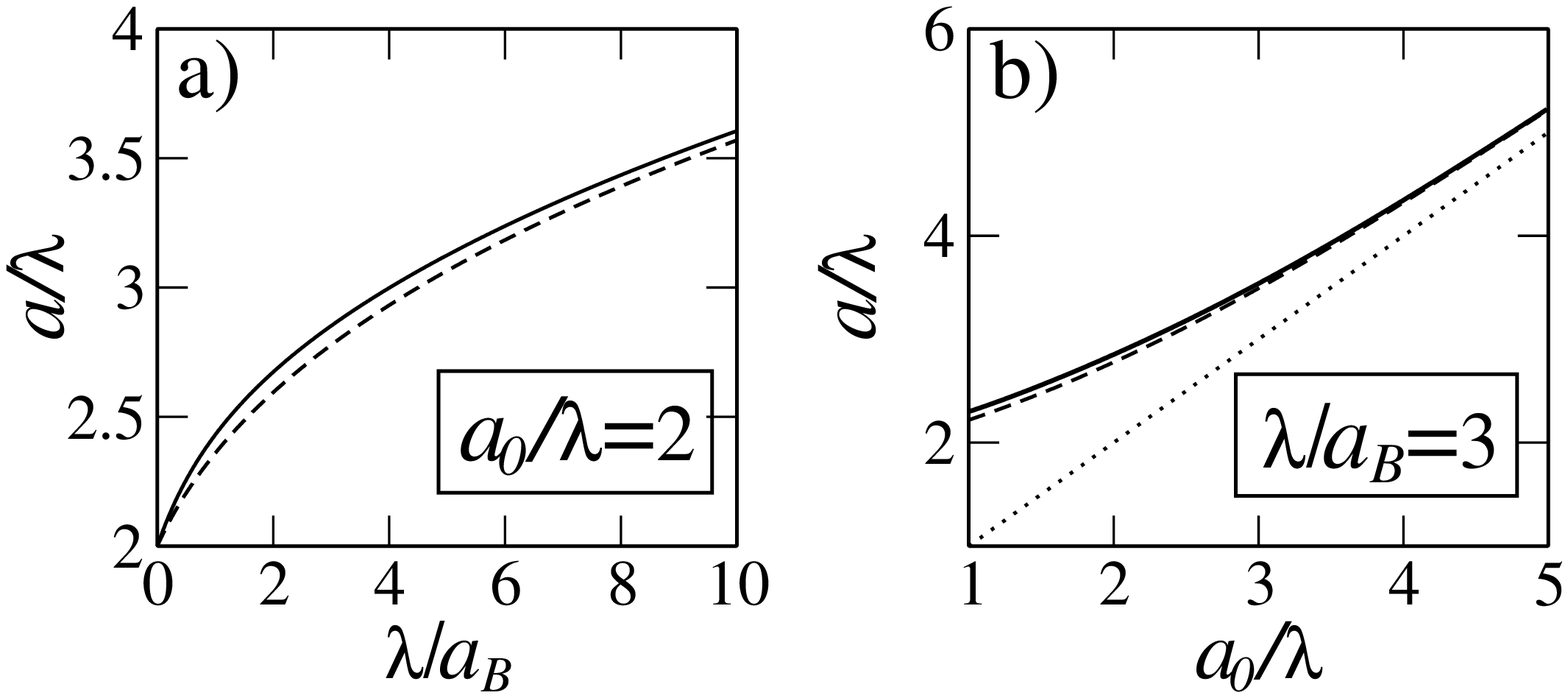}}
\caption{$a)$ The effective distance $a$ as a function of the Coulomb interaction strength $\lambda/a_B$ for $a_0/\lambda=2$. The full line represents the variational result from Eq. ($\ref{23}$). The dashed line corresponds to the one obtained from the classical equilibrium solution of  Eq. ($\ref{19}$). $b)$ Effective distance $a/\lambda$  as a function of the geometrical one $a_0/\lambda$ for $\lambda/a_B=3$. All distances are scaled with the dot radius $\lambda$. The dotted line is $a_0/\lambda$ which is 
shown for comparison.}
\label{Fig3}
\end{figure}with the definition $\omega_{x,y}=b_{x,y}\omega_0$ and $b_{x,y}$ given by the expressions
\begin{equation}
b_x=\sqrt{1+4(\lambda/a_B)(\lambda/a)^3},\quad b_y=\sqrt{1-2(\lambda/a_B)(\lambda/a)^3}.\label{22}
\end{equation}
We see that this approximation leads only to a renormalization of the orbital frequencies of the relative Hamiltonian. We proceed now to develop a second alternative for treating the Coulomb interaction, namely a variational method  based on the same picture of classical equilibrium. This consists in substituting the full Coulomb interaction term with the same type of expression like in Eq. ($\ref{19}$), but with the effective distance obtained from the variational ansatz. For this we minimize the expectation value of the orbital Hamiltonian, $H_r+H_C$, in the ground state of the effective relative  Hamiltonian $\widetilde{H}_r$[Eq. ($\ref{21}$)] with respect to the effective distance $a$. This leads to the following equation 
\begin{equation}
\frac{\partial}{\partial{a}}\langle\widetilde{\psi}_0|H_r+H_C|\widetilde{\psi}_0\rangle=0, \label{23}
\end{equation}
\begin{figure}[t]
\scalebox{0.27}{\includegraphics{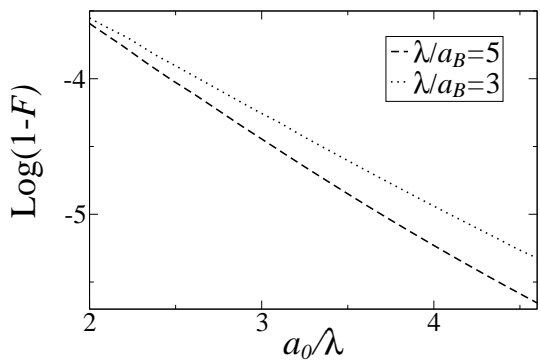}}
\caption{The logarithm of infidelity $1-F$ from Eq. ($\ref{28}$) as a function of the geometric distance $a_0/\lambda$ (scaled with the dot radius $\lambda$) for two different values of Coulomb interaction strength $\lambda/a_B$.}
\label{fidelity}
\end{figure}where $|\widetilde{\psi}_0\rangle$ is the ground state belonging to $\widetilde{H}_r$, $i.e.$ $\widetilde{H}_r|\widetilde{\psi}_0\rangle=\widetilde{E}_0|\widetilde{\psi}_0\rangle$, with $\widetilde{E}_0$ the ground state energy. Since we are dealing with harmonic oscillators, those wave functions are  known. However, Eq. ($\ref{23}$) for the effective distance $a$ can be solved only numerically. We plot in Fig. 3 the results obtained for the effective distance  $a$ as a function of different parameters  in both cases, namely the variational result from Eq. ($\ref{23}$) and also the result obtained from the classical equilibrium condition in Eq. ($\ref{19}$). We  see in Fig. 3  that there is  very good agreement between the two approaches in a wide parameter range and moreover, that the effective distance within the variational approach is larger then the one obtained from classical equilibrium, that means a lower ground state energy. We note that a perpendicular magnetic field practically does not change the curves  in Fig. 3 (not shown) on a large range of magnetic field strengths ($0<\omega_c<3\omega_0$), which means that the effective distance is to a very good approximation independent of the applied magnetic field. In order to verify the accuracy of our variational method, we checked also  the numerical fidelity, defined as the overlap of the wave functions in the variational case with the exact (almost, in the sense of perturbation theory) wave function. Although the problem contains no small parameter, we can still define small matrix elements compared with level spacing in a numerical sense. For this we write the full relative Hamiltonian in the following way
\begin{equation}
H_r=\widetilde{H}_r+V,\label{24}
\end{equation}
with the  effective Coulomb interaction  $V$ (see Eq. (\ref{20})) expressed now also in terms of the new equilibrium coordinates
(introduced after Eq. (\ref{20})),
\begin{equation}
V=\frac{e^2}{\kappa|\bm{r}+\bm{a}|}-\frac{e^2}{2\kappa{a^3}}\left(3(\bm{n}_{a0}\cdot\bm{r})^2-r^2\right)+\frac{e^2\,x}{\kappa{a^2}}.\label{25}
\end{equation}
We show now that this term  leads to small matrix elements so that indeed
$|V_{0n}|\ll|\widetilde{E}_n-\widetilde{E}_0|$, where the energies $\widetilde{E}_n$ and $\widetilde{E}_0$ are the n-eigenvalue and ground-state energy  of the Hamiltonian $\widetilde{H}_r$, respectively. To see this numerically we introduce the fidelity 
\begin{equation}
F=\left|\langle\psi_0|\widetilde{\psi}_0\rangle\right|^2,\label{26}
\end{equation}
where $|\psi_0\rangle$ and $|\widetilde{\psi}_0\rangle$ are the ground state wave functions of the full Hamiltonian $H_r$ and $\widetilde{H}_r$, respectively. We now estimate the fidelity $F$ by using  perturbation theory to find the true ground-state wave function $|\psi_0\rangle$ from the effective one $|\widetilde{\psi}_0\rangle$
\begin{equation}
|\psi_0\rangle=|\widetilde{\psi}_0\rangle+\sum_{n=1}^{\infty}\frac{\langle\widetilde{\psi}_n|V|\widetilde{\psi}_0\rangle}{\widetilde{E}_n-\widetilde{E}_0}|\widetilde{\psi}_n\rangle+\dots,\label{27}
\end{equation}
where we retain only terms to first order in $V$. Taking into account Eq. ($\ref{27}$) we  obtain the infidelity, $1-F$, namely the deviation of the true ground state wave function from the effective one
\begin{equation}
1-F=\sum_{n=1}^{\infty}\left|\frac{\langle\widetilde{\psi}_n|V|\widetilde{\psi}_0\rangle}{\widetilde{E}_n-\widetilde{E}_0}\right|^2.\label{28}
\end{equation}
We plot in Fig. 4  the infidelity $1-F$ as a function of the effective distance $a$ for  fixed Coulomb strength, $\lambda/a_B$. We see that the infidelity takes very small values ($1-F<10^{-2}$) on the considered range, for two different Coulomb strengths, which shows that our variational approach is very accurate.      
\begin{figure}[t]
\scalebox{0.55}{\includegraphics{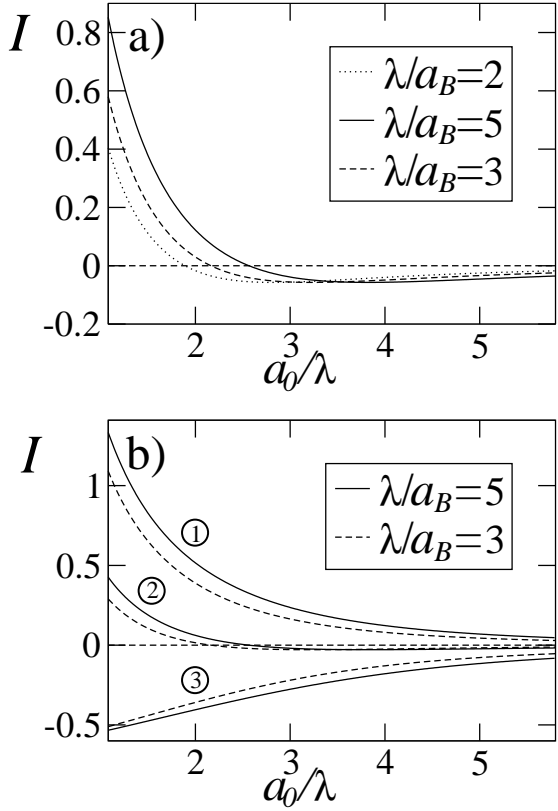}}
\caption{The function $I$  from Eq. ($\ref{30}$) as a function of the dimensionless geometric  distance $a_0/\lambda$. $a)$ The case of  perpendicular magnetic ($\theta=0$) field for  three different Coulomb strength parameters $\lambda/a_B$. $b)$ The  case of in-plane magnetic field ($\theta=\pi/2$), for two values of the Coulomb strength $\lambda/a_B$, for three  angles between the inter-dot distance vector $\bm{a}_0$ and the magnetic field. The groups of lines 1, 2, and 3 correspond to $\Phi=0$, $\Phi=\pi/4$, and $\Phi=\pi/2$, respectively.} 
\label{Fig4}
\end{figure}
 
 We can now evaluate the spin-spin interaction within this approximation. Since we are now  dealing with harmonic potentials only,  the problem of finding $J_{\tilde{x},\tilde{y}}$ from Eq. ($\ref{12}$) becomes straightforward. However, in the derivation of the spin Hamiltonian we need again some relations, similar to Eq. ($\ref{50}$) and  Eq. ($\ref{51}$), but for the present case with the harmonic oscillator renormalized. These relations read
\begin{equation}
L_R^{-1}X=-\frac{i}{\hbar M\omega_0^2}\left(P_X+\frac{eB_z}{c}Y\right)\label{60},
\end{equation}
\begin{equation}
L_R^{-1}Y=-\frac{i}{\hbar M\omega_0^2}\left(P_Y-\frac{eB_z}{c}X\right)\label{601},\nonumber
\end{equation}
\begin{equation}
L_r^{-1}x=-\frac{i}{\hbar m\omega_x^2}\left(p_x+\frac{eB_z}{c}y\right).\label{61}
\end{equation}
\begin{equation}
L_r^{-1}y=-\frac{i}{\hbar m\omega_y^2}\left(p_y-\frac{eB_z}{c}x\right).\label{611}\nonumber
\end{equation}
Making use of the relations Eq. ($\ref{60}$) and Eq. ($\ref{61}$), and also with the effective spin-orbit interaction expressed in the center-of-mass and relative coordinates, see Eq. ($\ref{8}$), the  spin-Hamiltonian $H_s$ takes the form
\begin{equation}
H_s=\frac{E_Z^2}{m^{*2}\omega_0^2\lambda_{SO}^2}\left[\left(1-\frac{1}{b_x^2}\right)\sigma_x^1\sigma_x^2+\left(1-\frac{1}{b_y^2}\right)\sigma_y^1\sigma_y^2\right],\label{62}
\end{equation}
for the case of a perpendicular magnetic field and 
\begin{equation}
H_s=\frac{E_Z^2}{m^{*2}\omega_0^2\lambda_{SO}^2}\left(1-\frac{\cos^2{\Phi}}{b_x^2}-\frac{\sin^2{\Phi}}{b_y^2}\right)\sigma_{\tilde{y}}^1\sigma_{\tilde{y}}^2,\label{63}
\end{equation}
for the case of an in-plane magnetic field which makes an angle $\Phi$ with the inter-dot distance direction. The $\tilde{y}$ is along the in-plane  direction perpendicular to the in-plane magnetic field. We see that  the spin Hamiltonian depends on the Coulomb interaction part via the difference between the inverse of the renormalized frequencies $\omega_{x,y}$ and the bare one $\omega_0$. As expected, when there is no renormalizations of the bare frequencies (no Coulomb interaction) the interaction vanishes. Referring again to the effective spin Hamiltonian $H^{{\it eff}}_s$ from Eq. ($\ref{13}$), we obtain for the coupling $J_{{\it eff}}$ for arbitrary magnetic field orientations
\begin{equation}
J_{{\it eff}}  =  E_z\;\frac{E_z}{\hbar\omega_0}\left(\frac{\lambda}{\lambda_{SO}}\right)^2\;I(a_0/\lambda,a_B/\lambda), \label{29} 
\end{equation}
where
\begin{eqnarray}
I(a_0/\lambda,a_B/\lambda)=\left(1-\frac{1}{b_x^2}\right)\left(\cos^2{\theta}\cos^2{\Phi}+\sin^2{\Phi}\right)\nonumber\\+\left(1-\frac{1}{b_y^2}\right)\left(\cos^2{\theta}\sin^2{\Phi}+\cos^2{\Phi}\right). \;\;\;\;\;\;\;\;\;\label{30}
\end{eqnarray}
 
%\begin{eqnarray}
%I(a_0,a_B)=\left\{
%\begin{array}{ll}
%\!\displaystyle{\frac{1}{b_x^2}+\frac{1}{b_y^2}-2,\;\;{\rm for\; B\parallel z}}\\
%\displaystyle{\frac{\cos^2{\Phi}}{b_x^2}+\frac{\sin^2{\Phi}}{b_y^2}-1,\;\;{\rm  for\; B\perp z}}.
%\end{array}
%\right.
%\label{30}
%\end{eqnarray}
One can see from Fig. 5b that for in-plane magnetic fields one obtains quite large values for $I$ in the two limiting cases $\Phi=0$ and $\Phi=\pi/2$. Changing the magnetic field orientation in-plane one can tune  the coupling strength $J_{{\it eff}}$ from negative to positive values, $i.e.$ from ferromagnetic to antiferromagnetic regime, and make it vanish for the angle (for in-plane magnetic field)
\begin{equation}
\Phi=\arcsin{\left(b_y\sqrt{(1-b_x^2)/(b_y^2-b_x^2)}\right)}\label{31}.
\end{equation}
In the case of a perpendicular magnetic field, cf.  Fig. 5a, we see that the coupling shows a non-monotonic behavior 
as a function of distance $a_0$, and, moreover, $J_{{\it eff}}$ vanishes for some given distance, 
which for $\lambda/a_B=5$ is about $a_0/\lambda\approx2.5$. 
This could be used to tune $J_{{\it eff}}$ on and off by changing the distance between the dots. 

Next, we consider the case of very elliptic dots, with the bare oscillator frequencies $\omega_{0x,0y}$ corresponding 
to the $x$ and $y$ directions, respectively, such that $\omega_{0x}\ll\omega_{0y}$. 
The spin-spin coupling becomes in this limiting case
\begin{equation}
H_s=\frac{E_Z^2}{m^{*2}\lambda_{SO}^2\tilde{\omega}_x^2}
\left(1-\frac{1}{b_x^2}\right)\cos^2{\Phi}\,
\sigma_x^1\sigma_x^2+O\big((\omega_{0x}/\omega_{0y})^2\big),
\end{equation}
where both perpendicular ($\Phi=0$) and in-plane magnetic fields are contained. 
We see that the problem becomes effectively 1D with an Ising-type spin-spin coupling, 
similar to the case of small Coulomb coupling studied in the previous section. 
We mention that for finite ratio of the two bare frequencies, $\omega_{0x}/\omega_{0y}$,
the interaction can be varied by changing this ratio, 
the angle of cancellation defined in Eq. ($\ref{31}$) varying as well.    

The behavior displayed in Fig. 5 can be understood as follows. The spin-spin coupling is directly 
related to the  deformation of the  charge distributions in the two dots as a consequence of 
the strong Coulomb interaction ($\delta\gg 1$). 
Thus, the stronger the deformation is, the stronger the spin-spin coupling becomes. 
Or, in our case, the stronger the deviation of the renormalized orbital frequencies $\omega_{x,y}$ from the bare one 
$\omega_0$ is, the stronger the coupling becomes, see Eqs. ($\ref{62}$) and ($\ref{63}$). 
While the $x$ component of the spin-spin coupling is bounded because the inverse of $\omega_x$ tends to zero as the 
Coulomb interaction strength $\delta$ increases, the $y$ component of this coupling is unbounded since the inverse of 
$\omega_y$ can grow indefinitely. 
Consequently, the $y$ component will dominate the $x$ component for large Coulomb strength and small inter-dot distance $a_0$. 
However, the situation is reversed in the large distance limit, since $\omega_x$ increases faster than $\omega_y$ decreases 
as seen from  Eq. ($\ref{22}$). These opposite limits lead to  the non-monotonic behavior depicted in Fig. 5a 
(perpendicular magnetic field). 

We mention that the large distance limit of Eq.~($\ref{29}$) converges to the large distance result obtained in the previous 
section, Eq.~($\ref{66}$). 
However, it does not converge to the results of the previous section in the case of small distance [Eq.~($\ref{16}$)], 
when going from $\delta\gg 1$ to $\delta\ll 1$, a crossover description ($\delta\sim 1$) being needed in this situation. 
Phrased differently, tuning the spin-spin coupling $J_{{\it eff}}$ from strong ($\delta\gg 1$) to small ($\delta\ll 1$) 
Coulomb interaction regime by varying the inter-dot distance reproduces the corresponding $\delta\ll 1$ result in 
Eq.~($\ref{66}$), while by varying the ratio $\lambda/a_B$ does not reproduce the corresponding $\delta\ll 1 $ 
limit, $i.e.$ Eq. ($\ref{16}$).        

Let us give now some estimates for the coupling $J_{{\it eff}}$ when an in-plane magnetic field is applied along, say,
the $x$-direction. 
Assuming now GaAs quantum dots, and  $E_Z=0.1\; {\rm meV}$ ($B=4 {\rm \; T}$), $\hbar\omega_0=0.5\; {\rm meV}$ 
($\lambda/a_B\approx5$), $\lambda/\lambda_{SO}\approx 10^{-1}$. 
Using these numbers and taking for the geometric inter-dot distance $a_0/\lambda\approx2$, we obtain 
$J_{{\it eff}}\approx10^{-7}\; {\rm eV}$. 
It is worth mentioning that the hyperfine interaction between the electron and the collection of nuclei in a quantum dot 
($\approx 10^5$) leads to similar energy scales\cite{KLG,CL1}. 
This shows that the spin-spin coupling derived here can be very relevant for the spin dynamics in the case of 
electrostatically coupled quantum dots and that it can also compete with other types of interactions. 
Considering now the case of InAs quantum dots in a magnetic field along the $x$ direction, with 
$\lambda_{SO}\approx 2\lambda\approx 100 {\rm nm}$ and $E_Z/\hbar\omega_0=0.1$ and taking also $a_0/\lambda\approx2$, 
a value of $J_{{\it eff}}\approx 10^{-6}{\rm eV}$ it is obtained. 
However, this is just a rough estimate since the spin-orbit coupling cannot be treated as a perturbation anymore 
and our approximation, being pushed to the limit of its range of validity, starts to break down. 

\section{Measurement Scheme}
\label{measurementschemelabel}
\begin{figure}
\scalebox{0.5}{\includegraphics{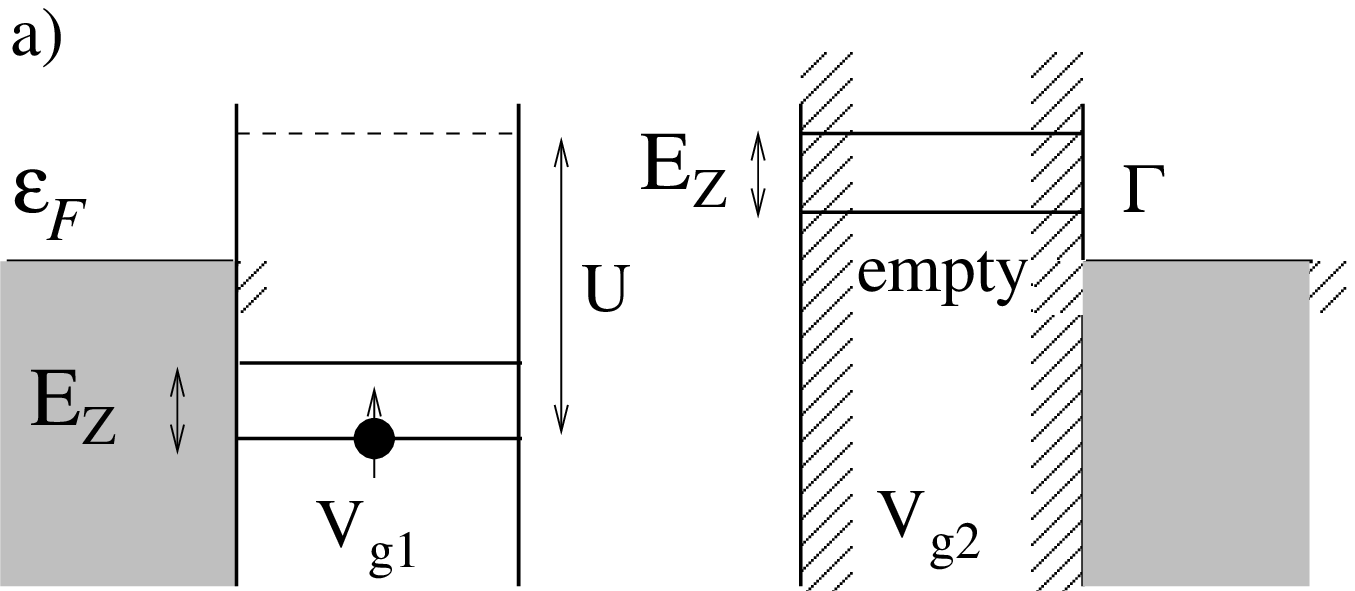}}\vspace{0.5cm}
\scalebox{0.5}{\includegraphics{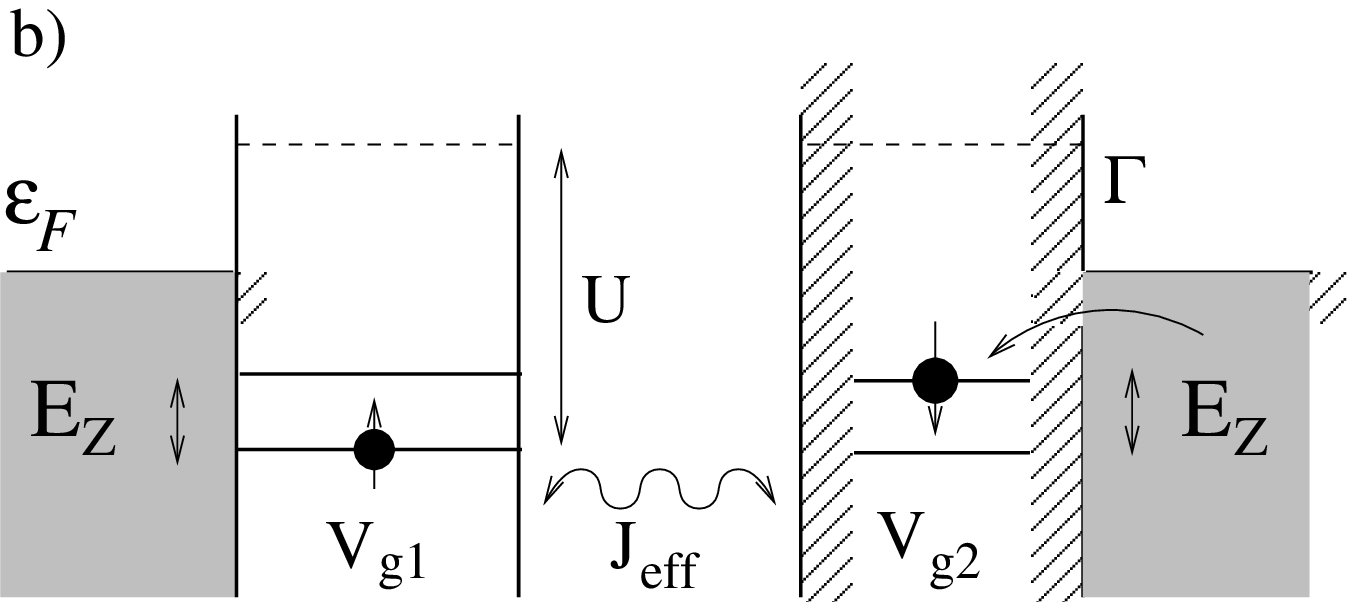}}\vspace{0.5cm}
\scalebox{0.5}{\includegraphics{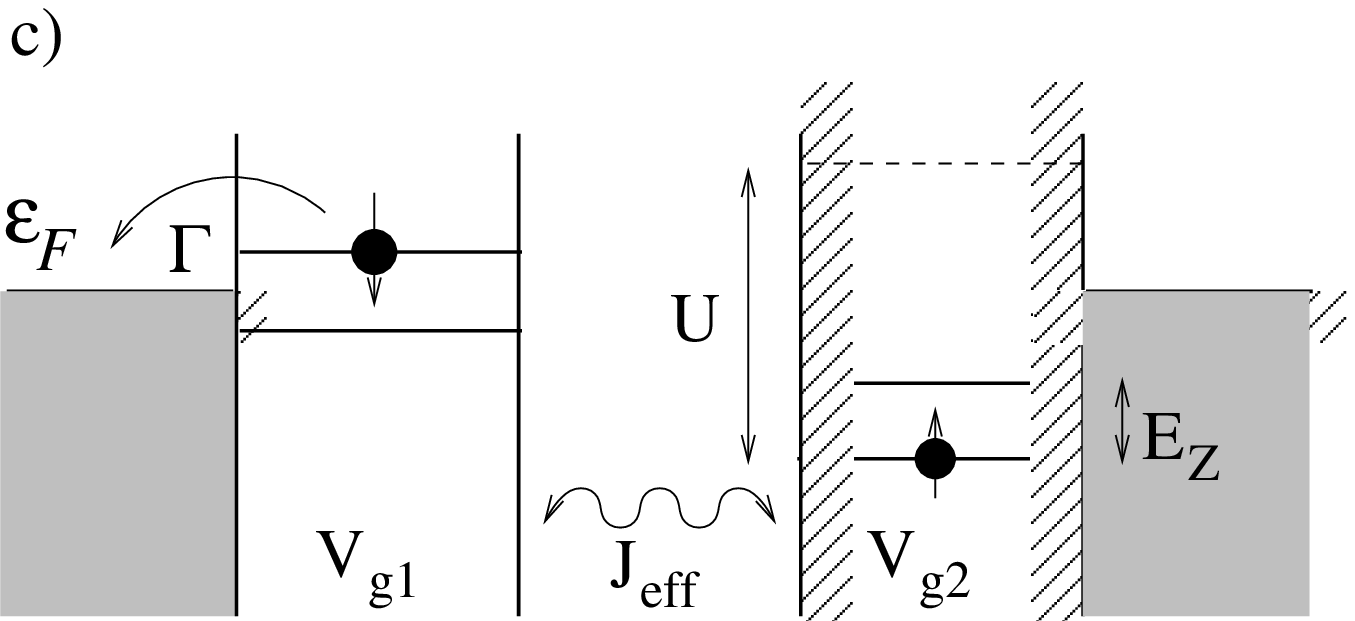}}
\caption{\small
Scheme to measure the coupling constant $J_{\it eff}$ in quantum dots without
tunnel coupling.
In the initialization step (a), the left dot is at equilibrium with one electron
in the lower Zeeman sublevel and the right dot is empty.
At the start of the coherent evolution step (b), the right dot is filled with one electron
in either upper or lower Zeeman level during a short time $\hbar/\Gamma\ll \hbar/J_{\it eff}$,
and the dots are deep in the Coulomb blockade valley, $\Gamma^2/U_\pm\ll J_{\it eff}$.
Further, the two spins evolve coherently due to the spin-spin interaction $J_{\it eff}$
during a fixed time $\tau\gtrsim \hbar/J_{\it eff}$.
In the read-out step (c), the left dot is brought up to the
Fermi surface, so that the electron can tunnel to the lead
only if it is in the upper Zeeman sublevel.
The latter event is recorded by a charge detector nearby the left dot.
}
\label{Schemefigure}
\end{figure}

In this section we propose a measurement scheme for the spin-spin interaction $J_{\it eff}$.
Similar to the spin relaxation experiments in Ref.~\onlinecite{Elzerman},
the left dot is monitored by a sensitive charge detector, such as
a quantum point contact (QPC) or a single electron transistor (SET).
We show the main steps of the scheme in Fig.~\ref{Schemefigure}. 

The first step is the initialization step shown in Fig.~\ref{Schemefigure}(a).
At low temperatures, $T\ll E_Z$, a single-electron dot will relax to the
ground state after a time larger than the spin relaxation time 
$T_1\simeq (0.1-100)\;{\rm ms}$. 
A faster spin relaxation can be induced by cotunneling with the lead, for
which the dot can be placed closer to the Fermi surface for some time.
In Fig.~\ref{Schemefigure}(a), the left dot is initialized in the
lower Zeeman sublevel $|\uparrow\rangle$, whereas the right dot is empty.
Next, the right-dot energy is lowered below the Fermi energy and the
dot is quickly filled with an electron in either upper or lower Zeeman sublevel.
This is a sequential tunneling process and we denote its rate by $\Gamma$.
In Fig.~\ref{Schemefigure}(b), both dots are deep below the Fermi surface
and $J_{\it eff}$ is the energy scale that governs a coherent evolution
in the subspace $\{|\uparrow\downarrow\rangle,|\downarrow\uparrow\rangle\}$.
The two spin density matrix reads
\begin{equation}
\varrho(t)=\frac{1}{2}|\uparrow\uparrow\rangle\langle\uparrow\uparrow|+
\frac{1}{2}|\Psi(t)\rangle\langle\Psi(t)|,
\end{equation}
where $|\Psi(t)\rangle$ is the wave function that describes the
occurrence of the state $|\uparrow\downarrow\rangle$ in the initialization step.
In the ideal case, $|\Psi(t)\rangle$ evolves coherently due to the
spin-spin interaction, Eq.~(\ref{13})
\begin{equation}
\Psi(t)\rangle=\cos(4J_{\it eff}t/\hbar)|\uparrow\downarrow\rangle-
i\sin(4J_{\it eff}t/\hbar)|\downarrow\uparrow\rangle.
\end{equation}
Here, we neglect the cotunneling and other spin relaxation processes.
In particular, the cotunneling rate $\Gamma^2/U_\pm$ should be
much smaller than the spin-spin coupling $J_{\it eff}$.
Here, $U_\pm$ is the addition/extraction energy of the single-electron
quantum dot in the Coulomb blockade valley.
On the other hand the sequential tunneling rate $\Gamma$ should
be large enough, so that the spins have no time to
evolve during the initialization and measurement steps.
We summarize the required regime by the inequality
\begin{equation}
\frac{\Gamma^2}{U_\pm}\ll J_{\it eff}\ll\Gamma.
\end{equation}
After a waiting time $\tau$, the probability of the left-dot electron
to be in the upper Zeeman sublevel reads,
\begin{equation}
P_{L\downarrow}(\tau)=\frac{1}{4}\left[
1-\cos(8J_{\it eff}\tau/\hbar)\right].
\end{equation}
Form the period of this function ($\tau_0=\pi\hbar/4 J_{\it eff}$) 
one can extract the value of the coupling constant $J_{\it eff}$.
%%%%%%%%%%%%%%%%%%%%%%%%%%%%%%%%%%%%%%%%%%%%%%
%           FIGURE
%%%%%%%%%%%%%%%%%%%%%%%%%%%%%%%%%%%%%%%%%%%%%%
%%%%%%%%%%%%%%%%%%%%%%%%%%%%%%%%%%%%%%%%%%%%%%

The measurement of the probability $P_{L\downarrow}(\tau)$ can be
performed in the same fashion as in Ref.~\onlinecite{Elzerman}.
After the waiting time $\tau$, the left dot is brought up to the 
Fermi level and placed such that the electron can tunnel into the lead
only from the upper Zeeman sublevel.
This configuration is shown in Fig.~\ref{Schemefigure}(c).
Tunneling of the electron out and refilling the quantum dot with an electron
of the opposite spin is monitored by the charge detector close to the
left dot (not shown).
For each value of the waiting time $\tau$, the cycle of initialization, coherent
evolution, and measurement has to be repeated many times in order
to reach a good accuracy.

Next, we remark that the hyperfine interaction with the lattice nuclei 
should not impede the measurement of $J_{\it eff}$ as long as 
$4J_{\it eff}\gtrsim A/\sqrt{N}$, where $A$ is the atomic hyperfine
coupling constant and $N$ is the number of nuclei in both quantum dots.
Note that the ratio of $J_{\it eff}$ to $A/\sqrt{N}$ for a constant $\lambda/a$ 
scales with the dot lateral size as $\propto\lambda^4$, for strong Coulomb interaction, 
and as $\propto\lambda^6$, for weak Coulomb interaction.
Therefore, the regime $4J_{\it eff}> A/\sqrt{N}$ can be easily achieved
by taking a larger quantum dot.
Furthermore, the hyperfine interaction with the nuclei has only the effect of 
reducing the visibility of oscillations of $P_{L\downarrow}(t)$,
and even for $A/\sqrt{N}\gg J_{\it eff}$ a small part of $P_{L\downarrow}(\tau)$
shows oscillatory behavior with unchanged period, $\tau_0=\pi\hbar/4 J_{\it eff}$.

\begin{figure}
\scalebox{0.7}{\includegraphics{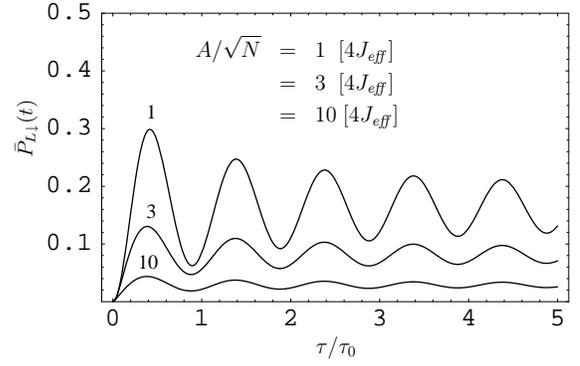}}
\caption{\small
Residual oscillations in the averaged probability $\bar{P}_{L\downarrow}(\tau)$
for values of $A/\sqrt{N}$ that exceed
the spin-spin interaction strength $4J_{\it eff}$.
The period of oscillations is not affected by the
hyperfine interaction and is given by $\tau_0=\pi\hbar/4 J_{\it eff}$.
With increasing the hyperfine strength $A/\sqrt{N}$, 
the amplitude of oscillations decreases as $\propto\sqrt N/A$.
As a function of the waiting time $\tau$, the envelope of
oscillations decays as $\propto 1/\sqrt\tau$.
}
\label{SchemefigureNucl}
\end{figure}

In Fig.~\ref{SchemefigureNucl}, we plot the probability $P_{L\downarrow}(\tau)$
averaged over the realizations of the hyperfine field. 
We choose $A/\sqrt{N} \geq 4 J_{\it eff}$ to show that
the measurement scheme is robust against the hyperfine field.
The oscillations are well visible even when $A/\sqrt{N}$ is several times
larger than $4 J_{\it eff}$.
The averaged probability $\bar P_{L\downarrow}(\tau)$ is obtained in the following way.
For the subspace $\{|\uparrow\downarrow\rangle,|\downarrow\uparrow\rangle\}$
and in the limit $E_Z\gg A/\sqrt{N}$,
the coupling of spins to the hyperfine field is given by
$H_{\delta h_z}={1\over 2}\delta h_z(\sigma_1^z-\sigma_2^z)$,
where the hyperfine field
$\delta h_z$ has a Gaussian distribution with zero average
and a variance $\sigma=A/\sqrt{N}$, which we take to be 
a measurable parameter that defines $N$.\cite{noteN}
For more detail on the derivation of $H_{\delta h_z}$ we refer the reader to
Ref.~\onlinecite{Klauser}.

For our description to be accurate, the time between subsequent cycles
of initialization, coherent evolution, and measurement
should be larger than the nuclear spin relaxation time.
Considering the sum of $H_{\delta h_z}$ and $H_s^{\it eff}$
in Eq.~(\ref{13}), we find that the probability
$P_{L\downarrow}(\tau)$ for a fixed value of $\delta h_z$
is given by\cite{note2} 
\begin{equation}
P_{L\downarrow}(\tau)=
\frac{1-\cos\left(2\tau\hbar^{-1}
\sqrt{16J_{\it eff}^2+\delta h_z^2}\right)}{4\left(1+\delta h_z^2/16 J_{\it eff}^2\right)}.
\label{PLdwndhz}
\end{equation}
The averaged probability $\bar P_{L\downarrow}(\tau)$ is then
computed by integrating Eq.~(\ref{PLdwndhz}) over $\delta h_z$
with the Gaussian weight factor
$P_\sigma(\delta h_z)={1\over\sigma\sqrt{2\pi}}e^{-\delta h_z^2/2\sigma^2}$,
where $\sigma=A/\sqrt{N}$.
%%%%%%%%%%%%%%%%%%%%%%%%%%%%%%%%%%%%%%%%%%%%%%
%           FIGURE
%%%%%%%%%%%%%%%%%%%%%%%%%%%%%%%%%%%%%%%%%%%%%%
%%%%%%%%%%%%%%%%%%%%%%%%%%%%%%%%%%%%%%%%%%%%%%

Considering $A/\sqrt{N}>4 J_{\it eff}$ and $\tau>\pi\hbar/4 J_{\it eff}$, we find that 
the visibility of oscillations scales with $A/\sqrt N$ and $\tau$ as follows
\begin{equation}
v\propto \frac{J_{\it eff}\sqrt N}{A}\sqrt{\frac{\hbar}{\tau J_{\it eff}}}.
\end{equation}
Note that the scaling law $v\propto \sqrt{N}/A$ is weaker than 
what one might expect na\"{i}vely from Eq.~(\ref{PLdwndhz}),
after substituting there $\delta h_z$ by its typical value $A/\sqrt{N}$, 
which gives $v\propto N/A^2$.
The reason for the weaker scaling law is the fact that $\delta h_z$ is centered around zero and
the denominator in Eq.~(\ref{PLdwndhz}) has nearly no effect.
We find that the numerical results in Fig.~\ref{SchemefigureNucl} can be reproduced fairly
accurately, if we approximate the argument of the cosine in Eq.~(\ref{PLdwndhz}) as follows
\begin{equation}
\sqrt{16J_{\it eff}^2+\delta h_z^2}\approx 4J_{\it eff}\left(1+\frac{\delta h_z^2}{32J_{\it eff}^2}+\dots\right).
\label{sqrtexpantion}
\end{equation}
This approximation is justified in the regime $\tau\gg\tau_0$ by the
the minimal phase requirement, despite the fact that $\delta h_z/J_{\it eff}$ may be large.
With Eq.~(\ref{sqrtexpantion}), it is easy to average $P_{L\downarrow}(\tau)$ 
and obtain an approximate expression, which is fairly accurate for $\tau\gtrsim\tau_0$ and asymptotically
exact in the limit $\tau\gg\tau_0$.
We thus obtain
\begin{eqnarray}
&&\bar{P}_{L\downarrow}(\tau)=\frac{1}{4}\left[\bar{p}-\delta p(\tau)\right],\label{Plppm1}\\
&&\bar{p}=\sqrt{\frac{\pi}{2}}\zeta \exp\left(\frac{\zeta^2}{2}\right){\rm erfc}
\left(\frac{\zeta}{\sqrt{2}}\right),\label{Plppm2}\\
&&\delta p(\tau)=\frac{\zeta\cos\left[2\pi\tau/\tau_0+
\varphi_0(\tau)\right]}{\left[\zeta^4+(2\pi\tau/\tau_0)^2\right]^{1/4}},
\label{Plppm3}
\;\;\;\;\
\end{eqnarray}
where $\zeta=4J_{\it eff}\sqrt{N}/A$, ${\rm erfc}(\zeta)$ is the complementary error function, and the
running phase shift $\varphi_0(\tau)$ is given by
\begin{equation}
\varphi_0(\tau)=\frac{1}{2}\arctan\left(\frac{2\pi\tau}{\tau_0\zeta^2}\right).
\label{varpho0Pl}
\end{equation}
We note that Eq.~(\ref{Plppm1}) is exact in two limiting cases: $\zeta\gg 1$ and $\tau\gg\tau_0$.
In particular, in the limit $\zeta\to \infty$, we recover
$P_{L\downarrow}(\tau)$ in Eq.~(\ref{PLdwndhz}) for all values of $\tau$.
In the opposite limit, $\zeta\ll 1$, Eqs.~(\ref{Plppm1})$-$(\ref{varpho0Pl}) can be significantly
simplified, yielding
\begin{equation}
\bar{P}_{L\downarrow}(\tau)=\frac{\zeta}{4}\left[\sqrt{\frac{\pi}{2}}-
\sqrt{\frac{\tau_0}{2\pi\tau}}\cos\left(2\pi\tau/\tau_0+\pi/4\right)\right].
\end{equation}

\section{Discussions and conclusions}
\label{discconclabel}

In the entire derivation we assumed  no tunneling  between the dots. Even in the presence of tunneling, when  direct Coulomb repulsion, $U_{12}$ (which is just the classical interaction between two charge distributions) is larger than the exchange interaction, $J_{exc}$ ($U_{12}\gg J_{exc}$), the theory presented here  is expected to remain still valid. The reason  is that this type of spin-spin coupling is a direct consequence  of the deformation of the  electronic charge distribution due to Coulomb repulsion between the two electrons. Since this is given by the sum of the direct Coulomb part $U_{12}$, and the exchange part $J_{{\it exc}}$, the spin coupling $J_{{\it eff}}$ will be insensitive to  exchange in the limit $U_{12}\gg J_{\it exc}$. The point now is that with finite tunneling, even with the assumption that the coupling strength $J_{{\it eff}}$ is not modified by the exchange Coulomb interaction, the resulting Heisenberg exchange  coupling  $J_{{\it exc}}$ ($H_{{\it exc}}=J_{{\it exc}}\bm{S}_1\cdot\bm{S}_2$)  
will start to compete with the electrostatically induced spin interaction. 
As a consequence, the spin coupling $J_{{\it eff}}$ will be washed out in the limit $J_{exc}\gg J_{{\it eff}}$. 
However, since the Heisenberg exchange coupling decays with the inter-dot distance  like\cite{BLV}  
$J_{{\it exc}}\sim \exp(-2a_0^2)$  while  $J_{{\it eff}}\sim a_0^{-3}$, the electrostatic  spin coupling  
will start to dominate at not very large  distances. 

We recall that when having exchange, another type of spin coupling, $J_{{\it exc}}^{SO}$, induced by spin-orbit interaction comes into play, and
which is proportional to the Heisenberg coupling $J_{{\it exc}}$, i.e. 
$J_{{\it exc}}^{SO}\propto (0.1-0.01) J_{{\it exc}}$ for typical GaAs  dots \cite{KVK,BSV,DBVBL,SB,BuL,BLV}. Thus,  $J_{{\it exc}}^{SO}$ can be  much larger than  
$J_{\it eff}$ for large enough Heisenberg  coupling, with a crossover from this exchange-type  
 to the 
direct Coulomb-type  coupling taking place at some inter-dot distance. 
This crossover, however,  occurs before we get  $J_{{\it exc}}\sim J_{{\it eff}}$, 
since $J_{{\it exc}}^{SO}$ is typically 100 times smaller than $J_{{\it exc}}$. 
To give an estimate, we assume $J_{\it exc}\approx 10^{-5} {\rm eV}$ for $a_0=1$, 
which gives $J_{\it exc}\approx3.5\cdot 10^{-8} {\rm eV}$ for $a_0=2$, implying that $J_{exc}<J_{{\it eff}}$.  

Another important  issue  is the effect of screening induced by the surrounding electrons in the 2DEG and the metallic gates. As is well-known, the screening effect between two charges  becomes important for distances exceeding the screening length $\lambda_{scr}\sim\lambda_{Fermi}$ (Fermi liquids). However, the screening of  bare Coulomb interaction depends strongly on the dimensionality. In 3D the effect of screening is to induce an exponential decay of  the bare Coulomb interaction \cite{PN}, with the decay parameter $\lambda_{scr}$, while in 2D  the decay follows a power law ($\sim r^{-3}$ in the large distance limit )  \cite{stern,ando,BH,SL},  with $\lambda_{scr}$ being the relevant length scale. For GaAs, the screening  length  is around $\lambda_{scr}\sim\lambda_{Fermi}\approx 50\, {\rm nm}$. Moreover, additional screening is introduced by the electrodes to gate the dots, due to their metallic character. The finite screening  implies then  that our theory in fact overestimates the strength of the electrostatically
induced spin coupling 
$J_{{\it eff}}$ for distances exceeding this screening length and the results obtained here become just an upper bound on $J_{{\it eff}}$ for this limit .

Being highly controllable, the  coupling $J_{{\it eff}}$ could  be used to perform two qubit gates for the realization of quantum computers  with electron spins, like proposed in Ref. \onlinecite{LDV}. The switching times range between rather slow ($\sim10 \,{\rm \mu s}$ in GaAs)
and reasonably fast ($\sim 50 \,{\rm ns}$ in InAs).
When making use  of the standard exchange coupling\cite{LDV} for switching 
(with typical switching times of
$ 100 \,{\rm ps}$ in GaAs) the electrostatically induced spin-coupling found here
can lead to gate errors. However, this effect can be controlled by choosing the magnetic field direction
or strength and/or the interdot distance such that $J_{{\it eff}}$ becomes negligibly small 
(see Eq. ($\ref{30}$)).

Finally, an  important question is how orbital fluctuations (for example of the confining energy  $\hbar\omega_0$) mediated via spin-orbit coupling lead to fluctuations in the coupling $J_{{\it eff}}$ and thus eventually to spin decoherence. The relation between the orbital dephasing time (which is assumed to be  known) and the decoherence induced by the spin coupling $J_{{\it eff}}$ reads \cite{CL}
\begin{equation}
\frac{\tau_{\phi}^o}{\tau_{\phi}^s}=\left|\frac{\delta{J}_{{\it eff}}}{\delta(\hbar\omega_0)}\right|^2, \label{32}
\end{equation}          
where $\tau_{\phi}^o$ is the orbital dephasing time and $\tau_{\phi}^s$ the corresponding spin decoherence time. We also assumed only fluctuations in the dot size, which means fluctuations of the confinement frequency $\omega_0$. Assuming an orbital dephasing time $\tau_{\phi}^o\approx 1\,{\rm ns}$ and also the limiting case of touching dots with the same GaAs parameters as before we obtain a spin decoherence time (lower bound) $\tau_{\phi}^s\approx 10^{-3}\,{\rm s}$. We can conclude then that the incoherent  part  due to this type of coupling is negligible compared with  other types of decoherence mechanisms, $e.g.$ induced by the hyperfine interaction\cite{KLG,CL1,CL}.
  
To conclude, we have derived an effective spin-spin interaction between two electrons localized in two quantum dots, spatially separated, induced by  the direct Coulomb interaction and mediated by the spin-orbit coupling. This interaction was found to have the form of an anisotropic $XY$ interaction and to be  proportional to the Zeeman energy. The spin-spin  coupling  was studied both in the weak and strong Coulomb interaction limits and for different magnetic field orientations and strengths. The important features are the non-monotonic behavior of this spin interaction for some magnetic field orientations, together with a vanishing of this interaction for particular inter-dot distances. This effect can be used to manipulate the spin-spin interaction in electrostatically coupled quantum dots by tuning the inter-dot distance. We proposed a measurement setup which allows one to access this spin-spin coupling experimentally.  

We thank D. Stepanenko, G. Burkard, W. A. Coish, and D. Klauser  for useful discussions.
We acknowledge financial support from the Swiss NSF, the NCCR Nanoscience, DARPA, ONR, and JST ICORP.\\
\\

\appendix

\section{$J_{\tilde{x},\tilde{y}}$ for arbitrary $\bm{B}$-fields}
%%%%%%%%%%%%%%%%%%%%%%%%%%%%%%%%%%%%%%%%%%%%%

In this Appendix we give explicit formulas for the couplings $J_{\tilde{x},\tilde{y}}$  for an arbitrary magnetic field orientation $\bm{B}=B(\cos{\Phi}\sin{\theta},\sin{\Phi}\sin{\theta},\cos{\theta})$ and for both Rashba and Dresselhaus spin-orbit couplings present. These are obtained by diagonalizing the tensor $\overline{\overline{M}}$,  which leads to 
%We first mention that we transform the frame in which the $z$ axis is perpendicular to the 2DEG plane and the $x$ axis is along the distance between the center of the two dots, $\bm{a}_0$ ($\bm{x}||\bm{a}_0$). The particular transformation consist in first rotating the $x$ and $y$ coordinates arround the $z$ axis such that $x$ lies along the projection of the magnetic field with the $y$ axis being perpendicular to the $\bm{B}$, followed by a rotation of the $z$ and $x$ axis arround the $y$ axis such that $z$ becomes parallel to $\bm{B}$. This rotation defines the $x''y''z''$ system of coordinates. The functions $A$, $B$ and $C$ become
%\begin{equation}
%A=f_1^2\,C_{y_1y_2}+f_1\,f_2\,(C_{x_1y_2}+C_{y_1x_2})+f_2^2\,C_{x_1x_2},
%\end{equation}  
%\begin{equation}
%B=f_3^2\,C_{y_1y_2}+f_3\,f_4\,(C_{x_1y_2}+C_{y_1x_2})+f_4^2\,C_{x_1x_2},
%\end{equation}
%\begin{equation}
%C=f_1\,f_3\,C_{y_1y_2}+f_1\,f_4\,C_{y_1x_2}+f_2\,f_3\,C_{x_1y_2}+f_2\,f_4\,C_{x_1x_2},
%\end{equation}
%where $C_{a_1b_2}=\langle 0|[L_d^{-1}a_1,b_2]|0\rangle$ ($a,b=x,y,z$) and 
%\begin{equation}
%f_1(\phi,\gamma)=\frac{\cos{\Phi}\cos{\gamma}}{\lambda_+}+\frac{\sin{\Phi}\sin{\gamma}}{\lambda_-},
%\end{equation}
%\begin{equation}
%f_2(\Phi,\gamma)=\frac{\cos{\Phi}\sin{\gamma}}{\lambda_+}-\frac{\sin{\Phi}\cos{\gamma}}{\lambda_-},
%\end{equation}
%\begin{equation}
%f_3(\Phi,\gamma)=\frac{\sin{\Phi}\cos{\gamma}}{\lambda_+}-\frac{\cos{\Phi}\sin{\gamma}}{\lambda_-},
%\end{equation}
%\begin{equation}
%f_4(\Phi,\gamma)=\frac{\sin{\Phi}\sin{\gamma}}{\lambda_+}+\frac{\cos{\Phi}\cos{\gamma}}{\lambda_-}.
%\end{equation} 
\begin{widetext}
\begin{eqnarray}
J_{\tilde{x},\tilde{y}} & = & \frac{1}{2}\big[(C_1+C_2)\cos^2{\theta}+(C_1\cos^2{\phi}+C_2\sin^2{\phi}-C_3\sin{2\phi})\sin^2{\theta}\nonumber\\ & \pm & \sqrt{\big((C_1+C_2)\cos^2{\theta}+\sin^2{\theta}(C_1\cos^2{\phi}+C_2\sin^2{\phi}-C_3\sin{2\phi})\big)^2-4(C_1C_2-C^2_3)\cos^2{\theta}}\big],
\end{eqnarray}
\end{widetext} 
with $\phi=\Phi-\gamma$ (the angle $\gamma$ is defined after Eq. ($\ref{8}$)). The functions $C_i$ ($i=1,2,3$) can be expressed in terms  of  $C_{a_1b_2}=\langle 0|[L_d^{-1}a_1,b_2]|0\rangle$, $a,b=x,y$, $i.e.$ 
\begin{equation}
C_1=\frac{1}{\lambda_-^2}\left(\sin^2{\gamma}\,C_{x_1x_2}+\cos^2{\gamma}\,C_{y_1y_2}+\sin{2\gamma}\,C_{x_1y_2}\right)
\end{equation}    
\begin{equation}
C_2=\frac{1}{\lambda_+^2}\left(\cos^2{\gamma}\,C_{x_1x_2}+\sin^2{\gamma}\,C_{y_1y_2}-\sin{2\gamma}\,C_{x_1y_2}\right)
\end{equation}
\begin{equation}
C_3  =  \frac{1}{2\,\lambda_+\lambda_-}\big(\sin{2\gamma}\,(C_{x_1x_2}-C_{y_1y_2})-\cos{2\gamma}\,C_{x_1y_2}\big).
\end{equation}
These functions can be identified very easily from our formulas derived in the paper. For example, for the case considered in Eq. ($\ref{56}$) (weak Coulomb coupling regime) we get
\begin{equation}
C_{x_1x_2}=\frac{\Delta E_C^x}{m^{*2}\lambda^2\omega_0^4},\;\;\;C_{y_1y_2}=\frac{\Delta E_C^y}{m^{*2}\lambda^2\omega_0^4},\;\;\;\; C_{x_1y_2}=0.
\end{equation}


\begin{references}
\bibitem{note1}
To distinguish between the two mechanisms we refer to the coupling studied
in this work as 'spin-spin coupling' as opposed to the Heisenberg exchange
coupling.
\bibitem{noteN}
More rigorously, $\sigma=IA/\sqrt{2N_0}$, where $I^2=\langle I_z^2\rangle$ 
is the variance of the $z$-component of a single nuclear spin (assuming $\langle I_z\rangle=0$) 
and $1/N_0=v_0\int d^3r|\psi({\bm r})|^4$, with $\psi({\bm r})$ being the electron wave function 
and $v_0$ the unit cell volume per nuclear spin.
\end{references}
\end{document}